# MORSE-PI – Flexible and artefact-free image reconstruction for structural and functional QSM and other phase-critical imaging applications.

Barbara Dymerska[1], Oliver Josephs[1], Benjamin James[1], Vahid Malekian[1], Nadine N. Graedel[1], Martina F. Callaghan[1]

*[1]The Functional Imaging Laboratory, Department of Imaging Neuroscience, UCL Queen Square Institute of Neurology, University College London, UK*

**Running Title:** MORSE-PI

**Corresponding author:**

Barbara Dymerska
Department of Imaging Neuroscience,
UCL Queen Square Institute of Neurology,
University College London,
London, WC1N 3AR, UK
Email: b.dymerska@ucl.ac.uk
Phone: +44-20-344-84383

## Acknowledgements

This research was funded in whole, or in part, by The Wellcome Centre for Human Neuroimaging [203147/Z/16/Z] and The Discovery Research Platform for Naturalistic Neuroimaging funded by the Wellcome [226793/Z/22/Z]. For the purpose of Open Access, the author has applied a CC BY public copyright licence to any Author Accepted Manuscript version arising from this submission. We thank Christian Lambert and Charlotte Dore for sharing challenging structural imaging data from the qMAP-PD study.

**Abstract:**

Purpose:

Phase imaging applications such as QSM are highly sensitive to noise amplifications, phase singularities, and other artefacts, particularly in challenging scenarios such as ultra-high field (7T), under-sampled or single-echo acquisitions. We present a novel image reconstruction method, MORSE-PI, designed to produce high-SNR, artefact-free, and singularity-free phase images for both structural (SWI, QSM) and functional (fQSM) phase-based brain imaging.

Methods:

MORSE-PI extends our previous approach, MORSE, by introducing a Virtual Reference Coil (VRC). The VRC is constructed as a linear combination of coil sensitivity maps, with correlations enhanced between coil elements using the noise covariance matrix. Such a VRC ensures robust signal support across the entire brain and is used to correct phase offsets in the MORSE-derived coil sensitivity estimates, resulting in artefact-free, high SNR phase.

Results:

Compared to GRAPPA with ASPIRE phase correction, MORSE-PI demonstrates greater robustness to artefacts such as noise amplification and aliasing, and shows improved reproducibility in structural imaging at both 3T and 7T. Unlike ESPIRiT and GRAPPA combined with adaptive coil combination methods, MORSE-PI yields singularity-free phase maps. MORSE-PI enables high-SNR reconstructions even for the most challenging scenarios, such as single-echo EPI at 7T. Its efficient, containerised implementation using the Gadgetron framework supports deployment on the MRI scanner console during measurements.

Conclusion:

MORSE-PI offers a flexible and computationally efficient solution for generating high-SNR, artefact- and singularity-free phase images in both single- and multi-echo GRE and EPI acquisitions. This makes it particularly well-suited for structural and functional QSM, as well as other phase-based MRI applications. Its robustness and rapid computational time facilitate efficient deployment on scanners across field strengths.

# 1. Introduction

Magnetic Resonance Imaging (MRI) is highly sensitive to subtle variations in magnetic susceptibility across the human brain, particularly when using T2*-weighted acquisitions with gradient-echo (GRE) or echo-planar imaging (EPI) readouts, and at high magnetic field strengths. These techniques enable the study of cerebral vasculature[1] as well as regional variations in tissue myelin and iron — two major contributors to magnetic susceptibility contrast[2,3]. This contrast also underpins blood oxygenation level-dependent (BOLD) functional MRI, which detects changes in blood oxygenation via associated susceptibility variations[4].

With the increasing availability of high-field MRI systems (≥3T), there has been rapid growth in high-resolution, phase-based imaging techniques sensitive to the magnetic properties of tissue, such as Susceptibility-Weighted Imaging (SWI)[5], Quantitative Susceptibility Mapping (QSM)[6], and functional QSM (fQSM)[7]. The broader adoption of these methods has been facilitated by advances in acquisition efficiency, image reconstruction methods, and improved understanding of phase contrast mechanisms. It is now well established that susceptibility contrast not only reflects tissue composition but also microstructural orientation, as exploited by methods such as Susceptibility Tensor Imaging (STI)[8]. In parallel, emerging techniques like magnetic susceptibility source separation have provided increasing insight into sub-voxel distributions of key biological substrates such as iron and myelin[9,10].

A crucial requirement for these techniques is the availability of robust, artefact-free, high-SNR phase images from T2*-weighted acquisitions. Such data are typically acquired using multi-channel receiver coils, which enhance SNR and support scan acceleration via under-sampling (i.e., the use of parallel imaging). However, combining the phase data from multiple coils is non-trivial[11]. In addition to magnetic susceptibility-induced phase perturbations (i.e., $\Delta B_0$-related phase terms), each coil also encodes phase contributions from its own sensitivity profile, as well as system-related effects including frequency drifts, timing errors and eddy currents. Further complications arise from the inherent 0 to 2π range of unprocessed phase data, which introduces spatial discontinuities ("phase wraps") that must be removed using phase unwrapping algorithms.

Incorrect handling of multi-coil phase information—for example, by disregarding non-$\Delta B_0$-related phase terms—can significantly degrade SNR, impairing downstream processing steps and rendering SWI and QSM results less sensitive and potentially unusable[12,13]. It can also introduce artefacts such as phase singularities, i.e. open-ended fringe lines or phase wraps that terminate within tissue rather than forming closed-loop phase contours. These phase artefacts constrain the selection of QSM processing steps and may mimic pathological tissue features such as microbleeds[12,13].

Several solutions exist that perform well for GRE-based phase imaging, particularly at 3T and lower field strengths, as reviewed in detail elsewhere[11,13]. However, the challenges increase at ultra-high fields (≥7T), where the lack of a body coil (often used as a phase reference in SENSE-based reconstructions) and the steeper coil sensitivity profiles complicate the generation of high-SNR, whole-brain phase images. Denser array coils further exacerbate this issue by producing more spatially localised and variable sensitivity patterns.

A well-established reconstruction method for 7T phase imaging is ASPIRE, which is a fast online (i.e., on the MRI console) approach that typically produces high-quality phase images from multi-echo GRE data without requiring additional reference scans or a body coil[14]. However, ASPIRE requires at least two echoes, and its phase image quality strongly depends on the underlying reconstruction of individual coil images, which may be affected by artefacts from parallel imaging methods such as GRAPPA. These artefacts can propagate into the final combined phase image.

We previously introduced MORSE[15]: a robust and computationally efficient method for coil sensitivity estimation and regularised SENSE[16] image reconstruction. We have shown that MORSE performs well at high field strengths (3T and 7T) and is more SNR-efficient than GRAPPA[17], in agreement with literature[18]. MORSE has been shown to reliably produce high-quality, alias-free magnitude images across a range of acceleration factors (demonstrated up to factor 8) for both structural and functional imaging. However, while the magnitude images from MORSE are of high quality, the associated phase images can suffer from phase singularities, which limit their utility in phase-sensitive applications.

In this work, we propose an extension termed MORSE-PI (PI = Phase Imaging), which flexibly generates singularity- and artefact-free, high-SNR phase images from single- or multi-echo GRE and EPI data here presented. MORSE-PI supports high-quality functional and structural SWI/QSM applications at both 3T and 7T. MORSE-PI is in active use within our department across a wide variety of research protocols, with online reconstruction implemented via a Docker container leveraging the Gadgetron framework[19].

In the following sections, we detail the modifications introduced in MORSE-PI relative to MORSE, present results from particularly challenging use cases identified during deployment, and compare its performance against commonly used image reconstruction strategies at 3T and 7T in the context of phase imaging.

# 2. Theory

We have recently released the open-source MORSE algorithm, which jointly estimates relative coil sensitivities and regularisation terms for use in a regularised SENSE image reconstruction framework[17]. We provide below a short theoretical summary of the method and, in the next section, explain the motivation behind the MORSE-PI extension proposed here.

In our notation, upper case non-italic bold letters indicate tensors ($\mathbf{T} \in \mathbb{C}^{I \times J \times K}$), upper case italic bold letters indicate matrices ($\boldsymbol{M} \in \mathbb{C}^{J \times K}$), lower case italic bold letters indicate column vectors ($\boldsymbol{v} \in \mathbb{C}^{K}$), and italic upper and lower case non-bold letters indicate scalars ($s \in \mathbb{C}$). Subscripts are used to index into a tensor, matrix or vector (*e.g.*, $T_{i,j,k}, M_{j,k}, v_k$). A superscript $H$ denotes the conjugate transpose ($\boldsymbol{X}^H$).

## 2.1 MORSE method

MORSE estimates coil sensitivity maps from integrated or separate fully sampled reference data, after applying coil compression via Singular Value Decomposition (SVD). This compression step enables the selection of a subset of coils that provide non-zero magnitude support across the entire imaging field of view, such that $N_{\mathrm{ref}} < N_{\mathrm{coil}}$, where $N_{\mathrm{coil}}$ denotes the total number of physical coils in the array. This $N_{\mathrm{ref}}$ subset of virtual coils serves as the reference in the construction of voxel-wise outer-product matrices $\boldsymbol{E}_r^{\mathrm{v}} \in \mathbb{C}^{N_{\mathrm{coil}} \times N_{\mathrm{ref}}}$:

$$\boldsymbol{E}_r^{\mathrm{v}} = \boldsymbol{c}_r^{\mathrm{v}} (\tilde{\boldsymbol{c}}_r^{\mathrm{v}})^H \tag{1}$$

where $\boldsymbol{c}_r^{\mathrm{v}} \in \mathbb{C}^{N_{\mathrm{coil}}}$ is the vector of complex image values at position $r$ in the virtual coil space, denoted by superscript v, for all coils, and $\tilde{\boldsymbol{c}}_r^{\mathrm{v}} \in \mathbb{C}^{N_{\mathrm{ref}}}$ is the corresponding vector truncated to the selected coil subset, $N_{\mathrm{ref}}$. A tensor built from these voxel-wise outer-product matrices, $\mathbf{E}^{\mathrm{v}} \in \mathbb{C}^{N_{\mathrm{coil}} \times N_{\mathrm{ref}} \times N_{\mathrm{voxel}}}$, is subsequently spatially smoothed, e.g., with a 3D Gaussian kernel, $\boldsymbol{g}$, (vectorised over space):

$$\hat{\mathbf{E}}^{\mathrm{v}} = \mathbf{E}^{\mathrm{v}} * \boldsymbol{g} \tag{2}$$

This convolution can be performed via computationally-efficient multiplication in k-space. A second economy voxel-wise SVD is applied to the spatially-smoothed outer-product matrices, $\widehat{\boldsymbol{E}}_r^{\mathrm{v}} \in \mathbb{C}^{N_{\mathrm{coil}} \times N_{\mathrm{ref}}}$:

$$\widehat{\boldsymbol{E}}_r^{\mathrm{V}} = \boldsymbol{U}_r \boldsymbol{S}_r \boldsymbol{V}_r^H = \sum_{k=1}^{N_{ref}} \sigma_{r,k} \boldsymbol{u}_{r,k} \boldsymbol{v}_{r,k}{}^H \tag{3}$$

The voxel-wise decomposition in Equation (3) produces $N_{\mathrm{ref}}$ pairs, indexed by $k$, of right singular vectors $\boldsymbol{v}_{r,k} \in \mathbb{C}^{N_{\mathrm{ref}}}$ and left singular vectors $\boldsymbol{u}_{r,k} \in \mathbb{C}^{N_{\mathrm{coil}}}$, with singular values, $\sigma_{r,k} \in \mathbb{R}^+$. A subset of left singular vectors, ($N_{\mathrm{order}} \leq N_{\mathrm{ref}}$) with singular values substantially greater than zero, is used as "higher-order" sensitivity estimates accounting for situations such as chemical shifts, rapidly varying sensitivities (e.g., particularly at the periphery of the brain), or spatial aliasing in fully sampled reference data caused by insufficient field of view. The corresponding singular values, $\sigma_{r,k}$ with $k = 1 \ldots N_{\mathrm{order}}$, are used for regularisation during the unfolding step.

## 2.2 SVD phase ambiguity

The SVD in Equation (3) suffers from a phase ambiguity, which is detrimental to phase imaging. This arises because an arbitrary phase offset, $\theta_{r,0}$, common to all $N_{\mathrm{coil}}$ vector elements, can be added to create a new set of left and right singular vectors that still fulfil the SVD decomposition: with new $\boldsymbol{u}'_{r,k} = e^{i\theta_{r,0}} \boldsymbol{u}_{r,k}$ and $\boldsymbol{v}'_{r,k} = e^{i\theta_{r,0}} \boldsymbol{v}_{r,k}$ it is valid that:

$$\sigma_{r,k} \boldsymbol{u}'_{r,k} \left(\boldsymbol{v}'_{r,k}\right)^H = \sigma_{r,k} e^{i\theta_{r,0}} \boldsymbol{u}_{r,k} e^{-i\theta_{r,0}} \boldsymbol{v}_{r,k}{}^H = \sigma_{r,k} \boldsymbol{u}_{r,k} \boldsymbol{v}_{r,k}{}^H \tag{4}$$

Due to this phase ambiguity, SVD implementations in various programming languages can produce different outcomes. In MATLAB, for example, the SVD implementation uses the LAPACK libraries, which ensures deterministic result by defining $\boldsymbol{v}_{r,1}{}^H$ as real for any complex matrix (here $\widehat{\boldsymbol{E}}_r^{\mathrm{V}}$).

Therefore, using voxel-wise SVD for the estimation of coil sensitivities can yield results with ambiguous, spatially varying phase offsets common to all the $N_{\mathrm{coil}}$ sensitivity estimates. These can cause phase singularities in the sensitivity estimates where the phase is not well defined. These singularities propagate to the phase of the reconstructed complex image even in regions with high SNR (see Results section). Such singularities substantially limit options for phase processing and QSM calculation yielding sub-optimal results, and they can necessitate separate calibration data to account for the effect, adding scan time.

Below we describe MORSE-PI, an extension to MORSE, which provides singularity-free phase images, preserves the high-quality magnitude images, and does not impose any requirements on the acquisition strategy.

## 2.3 MORSE-PI with Virtual Reference Coil

To remove ambiguous phase offsets and singularities we adopt the concept of a Virtual Reference Coil (VRC)[20] with scalar phase matching at a seed voxel[21,22]. In de Zwart et al. and Hammond et al. [21,22] a global phase offset is estimated for each receiver coil at a seed voxel, or within small reference region, and removed. This eliminates gross phase differences between the coils. Following this scalar phase matching, a VRC can be formed by a linear combination of the phase signals of all the receiver coils[20]. As discussed in the review paper by Robinson et al.[11], the original VRC approach often leads to phase singularities, which can move around depending on the selection of the seed voxel or the weights used for the linear combination of the signal from the separate coils.

To improve the robustness of the VRC approach in MORSE-PI, we construct the VRC from smooth MORSE-derived coil sensitivities, rather than from individual coil images, having enhanced the inter-coil correlation, and hence the spatial homogeneity of the VRC, using the noise covariance matrix. MORSE-derived coil sensitivities have spatially smooth profiles. Situations in which seed voxels are placed in unreliably low signal regions (e.g., a vessel) that yield poor phase matching between the coils are avoided because the sensitivities are free of tissue contrast. Using the noise covariance matrix to ensure good correlation between coil sensitivities, prior to VRC formation, provides good VRC magnitude and phase support, typically over the entire brain without phase singularities, as explained below.

In many MRI applications, the noise covariance matrix is calculated from short noise measurements (without any excitation of MR signal) prior to running the imaging sequence, and used to decorrelate, or whiten, the complex signals from separate coils. This is typically achieved by factorising the noise covariance matrix, $\boldsymbol{\psi} \in \mathbb{C}^{N_{\text{coil}} \times N_{\text{coil}}}$, using Cholesky decomposition: $\boldsymbol{\psi} = \boldsymbol{L}\boldsymbol{L}^H$, where $\boldsymbol{L} \in \mathbb{C}^{N_{\text{coil}} \times N_{\text{coil}}}$ is a lower triangular matrix. The whitening is then performed by multiplying $\boldsymbol{L}^{-1}$ with the complex data. This yields whitened signals with identity noise covariance. This also leads to a new set of composite coil sensitivities which, besides decorrelated noise, are more spatially independent than the original set, i.e., have less spatial overlap between neighbouring coils. More independent noise between coils improves the SNR of the combined/unfolded image, and steeper coil sensitivities increase the unfolding efficacy. However, VRC construction requires sufficient overlap between neighbouring coil sensitivities to prevent signal voids and phase singularities in regions with poor sensitivity coverage. Therefore, having initially decorrelated the separate coil signals, i.e., whitened the data, we subsequently use the lower triangular matrix to enhance the correlation between the first order MORSE-derived coil sensitivities:

$$\widetilde{\boldsymbol{b}}_{r,1} = \boldsymbol{L}^n \widehat{\boldsymbol{b}}_{r,1} \tag{5}$$

where $\widehat{\boldsymbol{b}}_{r,1} \in \mathbb{C}^{N_{\text{coil}}}$ is a vector with the original first order coil sensitivity estimates in physical coil space at a given spatial location $r$. The lower triangular matrix $\boldsymbol{L}$ is raised to $n$-th power ≥ 1 to enhance the correlation between coil sensitivity profiles and ensure singularity-free VRC. $\widetilde{\boldsymbol{b}}_{r,1} \in \mathbb{C}^{N_{\text{coil}}}$ are the corresponding values of the correlated first order coil sensitivities. A simple complex sum over coils of $\widetilde{\boldsymbol{b}}_{r,1}$, after scalar phase matching at the image centroid, $r_c$ , creates the final VRC, the phase of which, $\varphi_r^{VRC}$, is used to correct the original first order coil sensitivities:

$$\varphi_r^{VRC} = \angle \left[ \sum_{k=1}^{N_{\text{coil}}} \tilde{b}_{r,1,k} \cdot e^{-i\varphi_{r_c,k}} \right] \tag{6}$$

$$\hat{b}_{r,1,k}^{MORSE-PI} = \hat{b}_{r,1,k} \cdot e^{-i\varphi_r^{VRC}} \tag{7}$$

The higher order coil sensitivities $\hat{B}_{r,2\ldots N_{\text{order}},k}$ are not modified since we empirically established that their contribution to the final reconstructed phase image is small, and does not cause phase singularities. The computation is presented above voxel-wise for clarity. The actual computation is done using computationally efficient matrix operations. The image centroid position $r_c$ is selected based on the root-sum-of-squares across coils of the original fully sampled image domain reference data used to estimate the sensitivities.

Note, it is essential to distinguish between the $N_{\text{ref}}$ reference coils used in constructing the outer-product matrices (as in Equation (1)) and the single VRC employed for the final phase correction (as in Equation (6)). While both serve as reference coil entities, they pertain to distinct stages of the image reconstruction process.

As in the original MORSE approach, we use a regularised SENSE formalism to reconstruct the final complex images, but now free from phase singularities. The final magnitude image is identical to that obtained with MORSE, i.e., the phase correction from Equation (7) does not influence the signal magnitude.

# 3. Methods

The MORSE-PI image reconstruction scheme has been deployed for a diverse range of 3T and 7T functional and quantitative structural imaging studies at The Functional Imaging Laboratory (The FIL), University College London. The open-source code, along with scripts for creating a Docker container that encapsulates the MORSE-PI reconstruction within the Gadgetron framework[19], is available on GitHub[23]. Below, we describe the implementation and summarize the performance evaluation carried out on representative *in vivo* data prior to deployment.

## 3.1 Gadgetron and MATLAB implementation

Figure 1 visualises the computational steps of the MORSE-PI phase correction applied to the first order coil sensitivities calculated using MORSE. The noise covariance matrix is computed using the native NoiseAdjustGadget in Gadgetron[24]. This noise covariance matrix is used to pre-whiten k-space data (fully sampled reference and accelerated data) and, independently, to correlate the estimated coil sensitivities for VRC creation.

As noted previously, only the phase of the first-order coil sensitivities estimated by MORSE was corrected using the VRC phase, following the workflow illustrated in Figure 1. The magnitude of the first-order coil sensitivities and the higher-order complex components were left unchanged from the original MORSE implementation. The MORSE-PI processing steps were integrated into the original MORSE algorithm in MATLAB (Mathworks, Natick, MA).

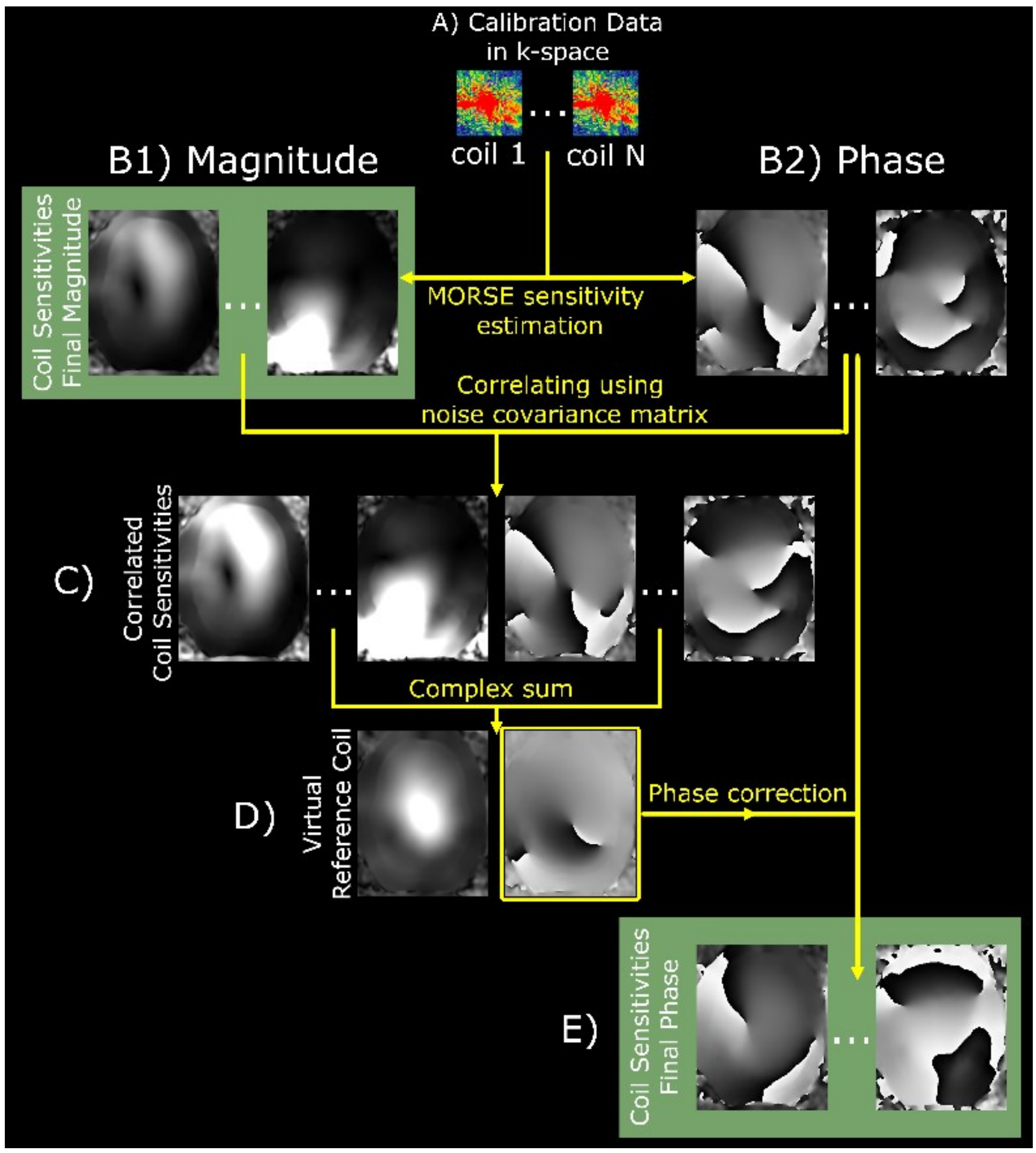

**Figure 1** Phase correction steps in MORSE-PI. Yellow text indicates processes, white text denotes data inputs/outputs, and green highlights the final first-order coil sensitivities (magnitude from step B1, and phase from step E. Complex coil sensitivities (B1 and B2) are estimated from fully sampled k-space reference data (A) and subsequently correlated using Equation (5), where $n = 1$ was empirically determined to be sufficient to yield singularity-free results in data here presented. The VRC sensitivity in step D is then computed as the complex sum of the correlated coil sensitivities, using scalar phase matching as described in Equation (6). This process ensures robust VRC support across the entire brain (see VRC magnitude image in D). The VRC phase (highlighted by the yellow box in D) is then applied for coil-wise phase correction resulting in the final first-order coil sensitivity phases in E. Residual phase singularities in E remain only in regions lacking magnitude support for the respective coil (see B1), and therefore do not propagate to the final phase image.

## 3.2 In vivo data acquisition

MORSE-PI was evaluated using data collected following ethical approval from local Research Ethics Committee and informed consent from study participants. The 3T data were acquired on a Siemens Healthineers Prisma$^{fit}$ (software version VE11c) using a body coil for transmission and a 64-channel head and neck array for reception. The 7T data were obtained on a Siemens Healthineers Terra (software version VE12u-SP01) with a Nova Medical head coil featuring 8 transmit and 32 receive channels, operated in a quadrature-like transmit mode. Table 1 summarizes the key protocol parameters for these acquisitions.

**Table 1** Sequence parameters for 3T and 7T GRE and EPI scans on which MORSE-PI reconstruction was evaluated.

<table>
<tr><td colspan="8">Protocol for Structural Imaging: 3D Gradient Echo</td></tr>
<tr><th>Field [T]</th><th>Resolution [$mm^3$]</th><th>Acceleration</th><th>Reference data</th><th>Matrix [HF X AP x RL]</th><th>TR [ms]</th><th>TE, ΔTE [ms]</th><th>TA [min:s]</th></tr>
<tr><td>7</td><td>0.6 x 0.6 x 0.6</td><td>2 (PE) x 2 (3D)</td><td>48 x 48 Integrated reference lines</td><td>426 x 364 x 288</td><td>19.5</td><td>2.3, 2.3</td><td>9:10</td></tr>
<tr><td colspan="8">Image quality assessment: PD or T1 weighting was achieved with a flip angle of 6° or 24° respectively; MT weighting was achieved with a 190° pre-pulse 2 kHz off-resonance and an excitation flip angle of 6°. The PD and T1 weighted acquisitions had 6 echoes while the MT-weighted acquisition had 4.<br>The PD-weighted protocol was additionally run with the vendor-provided sequence with minor modifications to the receiver bandwidth and TR, leading to a TA of 9:15.</td></tr>
<tr><td>7</td><td>1.0 x 1.0 x 1.0</td><td>2 (PE) x 2 (3D)</td><td>44x48 Integrated reference lines</td><td>256 x 224 x 176</td><td>19.5</td><td>2.3, 2.3</td><td>3:48</td></tr>
<tr><td colspan="8">Reproducibility assessment: The PD and T1 weighted acquisitions were repeated 3 times during the same scan session.</td></tr>
<tr><td>3</td><td>0.8 x 0.8 x 0.8</td><td>2 (PE) x 2 (3D)</td><td>40 x 40 Integrated reference lines</td><td>320 x 280 x 224</td><td>25.0</td><td>2.3, 2.3</td><td>7:08</td></tr>
<tr><td colspan="8">Image quality assessment: PD or T1 weighting was achieved with a flip angle of 6° or 21° respectively; MT weighting was achieved with a 220° pre-pulse 2 kHz off-resonance and an excitation flip angle of 6°. The PD and T1 weighted acquisitions had 8 echoes while the MT-weighted acquisition had 6.</td></tr>
<tr><td>3</td><td>1.0 x 1.0 x 1.0</td><td>2 (PE) x 2 (3D)</td><td>40 x 40 Integrated reference lines</td><td>256 x 224 x 176</td><td>25.0</td><td>2.3, 2.3</td><td>4:42</td></tr>
<tr><td colspan="8">Reproducibility assessment: The PD and T1 weighted acquisitions were repeated 3 times during the same scan session.</td></tr>
<tr><td colspan="8">Protocol for Functional Imaging: 3D EPI</td></tr>
<tr><th>Field [T]</th><th>Resolution [$mm^3$]</th><th>Acceleration</th><th>Reference data</th><th>Matrix [HF x AP x RL]</th><th>TR [ms]</th><th>TE [ms]</th><th>TA [s/vol]</th></tr>
<tr><td>7</td><td>0.8 x 0.8 x 0.8</td><td>None, 8 in-plane segments</td><td>None</td><td>Whole brain 160 x 240 x240</td><td>100.0</td><td>18.1</td><td>4:49</td></tr>
<tr><td>7</td><td>0.8 x 0.8 x 0.8</td><td>4 (PE) x 2 (3D)</td><td>Separate with matched resolution</td><td>Slab-selective 88 x 240 x 240</td><td>43.0</td><td>17.7</td><td>3.8</td></tr>
<tr><td colspan="8">Whole-brain acquisition (2 volumes) was run with MT pre-pulse (flip angle 300° 2 kHz off-resonance) for better grey-white matter contrast and improved coregistration of functional to structural data[25].<br>50 task-free volumes were run with a slab-selective protocol for temporal stability assessment (tSNR).</td></tr>
</table>

### 3.2.1 Impact of cross-coil correlation on VRC spatial support

The strength of the correlation between coil sensitivities can be adjusted by varying $n$ in Equation (5). We tested values of $n$ = -1, 0, 1, 2, 3 to evaluate the VRC magnitude support across the regions of interest, namely the brain and cervical spinal cord in 3T and 7T structural scans. We also assessed the presence or absence of phase singularities in the VRC. To examine noise correlation between neighbouring coils, we also applied the $\boldsymbol{L}^n$ operator to the noise covariance matrix itself, i.e. $\boldsymbol{L}^n \cdot \boldsymbol{\psi} \cdot (\boldsymbol{L}^n)^H$. This analysis began with $n$ =-1 yielding the decorrelated, i.e. whitened noise covariance matrix (${\boldsymbol{\psi}_{whitened}}^{n=-1} = \boldsymbol{L}^{-1} \cdot \boldsymbol{\psi} \cdot (\boldsymbol{L}^{-1})^H = \boldsymbol{I}$), proceeded to $n$ =0 with the inherent noise covariance matrix (i.e. ${\boldsymbol{\psi}_{inherent}}^{n=0} = \boldsymbol{\psi}$ ) - equivalent to no prewhitening, and extended up to $n$ =3, corresponding to the 3rd power of the correlation (${\boldsymbol{\psi}_{correlated}}^{n=3} = \boldsymbol{L}^3 \cdot \boldsymbol{\psi} \cdot (\boldsymbol{L}^{-3})^H$).

### 3.2.2 Structural imaging at 7T and 3T - comparison with alternative image reconstruction schemes

MORSE-PI was compared with the original MORSE implementation, as well as methods commonly used for phase imaging: vendor-provided GRAPPA[26] + Adaptive Combine[27] (with and without SVD) and state-of-the-art phase imaging with GRAPPA[26] + ASPIRE[14], and ESPIRiT[18]. The GRAPPA + Adaptive Combine (with and without SVD) and GRAPPA + ASPIRE images were reconstructed using their default image reconstruction chains (via the Siemens Image Calculation Environment) on the native scanner hardware via the "retrospective reconstruction" facility. For GRAPPA + Adaptive Combine with SVD, the image reconstruction instructions from the QSM Consensus Organization Committee publication Supporting Information III[13] were followed. The MORSE, MORSE-PI and ESPIRiT reconstructions were performed offline on a Dell Precision 7920 with Intel Xeon Gold 6230R 2.1GHz Processor and 768 GB of RAM. For ESPIRiT the BART toolbox v0.4.0.4 was used with and without coil compression prior to calling the "ecalib" function for coil sensitivity estimation and the "pics" function for unfolding. From our previously reported comparisons[17], we deemed ESPIRiT without regularisation and with the "Soft Sense" option (i.e. "ecalib -m 2") to be an appropriate choice for the structural imaging data assessed here. For all structural data, the MORSE and MORSE-PI reconstructions used: $N_{ref} = 6$, $N_{order} = 4$, gaussian smoothing filter $w = 6$ voxels, and global regularisation factor of $\lambda$ = 3x10$^{-4}$, as deployed in various *in vivo* studies at The FIL[17].

Structural images were acquired at 3T and 7T using an in-house, but openly available, 3D multi-echo spoiled gradient echo acquisition configured with either proton density (PD), T1 or magnetisation transfer (MT) weighting. An under-sampling factor of two was used in each phase-encoded direction, with integrated reference data. To further ease replication at other sites, additional data were acquired at 7T with a vendor-provided sequence. Full details of all acquisitions are provided in Table 1.

MR phase imaging is commonly used for SWI and QSM, we therefore prepared image processing pipelines for these contrasts and share them on GitHub[28]. These pipelines were employed to assess final image quality, as well as noise and error propagation from the image reconstructions used in this study. For SWI, we applied the state-of-the-art CLEAR-SWI[29] multi-echo method with default options of magnitude sensitivity correction (i.e. bias field removal) and Laplacian phase unwrapping, and two options adjusted to the resolution of our data: high-pass filter of size [2 2 0] voxels and Minimal Intensity Projection (MIP) across 14 axial slices (for depicting vessel networks).

The QSM pipeline included phase unwrapping and field map calculation using ROMEO[30], hybrid masking using ROMEO[30] and SPM[31] segmentation[32], background field removal via PDF[33], and dipole inversion using STAR-QSM[34]. We used CLEAR-SWI and ROMEO from the MriResearchTools package (v4.5.3), while PDF and STAR-QSM were called via the SEPIA[35] toolbox interface, which allowed fast testing of various QSM methods combinations. ROMEO uses weighted echo averaging for field map calculation[36], which is deemed the optimal solution for QSM if the echo times are chosen appropriately[13,37]. This solution can, however, lead to propagation of phase singularities present in MORSE, GRAPPA + Adaptive Combine, and ESPIRiT reconstructions to the final result. We thus also explored an alternative solution, namely a non-linear complex fit[38], prior to ROMEO phase unwrapping, to assess the impact of phase singularities.

Masks were created by binarizing the ROMEO quality map with a 0.1 threshold and multiplying the resulting mask with an SPM-based mask defined as those voxels that had an integrated grey matter, white matter and cerebrospinal fluid tissue probability greater than 0.95. This mask creation approach was applied consistently to all image reconstructions. Finally, to ensure a fair comparison, a common mask was generated by multiplying all the hybrid ROMEO-SPM masks derived from each of the different image reconstructions.

To minimise the required field of view while avoiding aliasing artefacts, all structural acquisitions were performed with an oblique orientation about the left-right axis. Consequently, the field map and brain mask were resliced such that one image axis was aligned with the main magnetic field ($B_0$) direction prior to background field removal. The final QSM maps were referenced by subtracting the mean susceptibility value within the whole brain mask.

To assess the reproducibility of the QSM maps (result section 4.2.2 Reproducibility of QSM reconstructions), three pairs of PD-weighted and T1-weighted scans with 1 mm isotropic resolution were acquired from the same volunteer during a single scan session at 3T. This was repeated at 7T with the same volunteer. To correct for inter-scan motion effects, motion estimates obtained by co-registering the magnitude data were used to realign the QSM maps. Subsequently, the mean and standard deviation (SD) of the QSM maps were computed, along with histograms of SD values. These metrics were evaluated for each contrast separately using the singularity-free reconstruction methods: GRAPPA+ASPIRE and MORSE-PI.

### 3.2.3 Functional imaging at 7T – evaluation of MORSE-PI phase quality and temporal stability

To assess MORSE-PI in a challenging scenario where no established alternative solutions for singularity-free phase imaging are currently available, we performed high-resolution (0.8 mm isotropic) 3D EPI scans at 7T where no reference body coil scan is available (yet required for original SENSE image reconstruction[39]). The acquisition included only single-echo data, as typically used for laminar fMRI studies[40], prohibiting the use of the ASPIRE coil combination.

We employed an established 3D EPI protocol in use in cognitive neuroimaging studies at the FIL, specifically: (1) a whole-brain Magnetisation Transfer (MT) weighted acquisition without acceleration, and (2) a slab-selective, accelerated time series acquisition typical of functional MRI studies[25]. The details of these acquisitions are listed in Table 1. The MORSE-PI reconstruction used: $N_{ref} = 8$, $N_{order} = 2$, gaussian smoothing filter $w = 6$ voxels, and global regularisation factor of $\lambda$ = 0.1.

For the EPI time series ROMEO phase unwrapping was used with the “individual unwrapping” option, where each EPI volume was unwrapped individually rather than using a template – this minimises the unwrapping errors in long time series. The unwrapped phase was converted to Hertz (i.e. divided by 2πTE). To assess the suitability of MORSE-PI reconstructed phase for various QSM image processing methods two alternative background field removal methods: PDF[33] and RESHARP[41] and two dipole inversion methods: STAR-QSM[34] and QSMnet+[42] were investigated. The mean of the QSM maps for the time series, and their temporal standard deviation (tSD), were calculated for three QSM processing pipelines: (PDF → STAR-QSM, PDF → QSMnet+, RESHARP → STARQSM).

# 4. Results

## 4.1 Impact of cross-coil correlation on VRC spatial support

Figure 2 shows the VRCs calculated using Equation (6) when noise correlation is unavailable (A, no pre-whitening) or after pre-whitening has been applied prior to estimating the coil sensitivities (B-D) for a 8 transmit and 32 receive channel coil at 7T. For the pre-whitened data, results are shown for coil sensitivities correlated across coils, according to Equation (5), with powers of $n = -1, 0, 1$ (B, C, and D corresponding to sensitivities that are whitened, have the original cross-coil correlation, or enhanced cross-coil correlation respectively). A phase singularity, marked by a green arrow, arises from the voxel-wise SVD performed during MORSE coil sensitivity estimation and is present in all VRCs. If no VRC phase correction is applied, this singularity propagates into the reconstructed phase image ($4^{th}$ column in Figure 2) but is removed by VRC phase correction ($5^{th}$ column in Figure 2). In Figure 2A and 2B, additional unwanted phase singularities (marked with red arrows) are present where there is very low VRC magnitude regardless of having applied the correction. In Figure 2C ($n = 0$), a residual high-spatial-frequency artefact remains in the VRC phase (red arrow in C), which propagates into the final phase image. However, after increasing the cross-coil correlation with $n = 1$ (Figure 2D), no region within the brain or cervical spine has VRC magnitude values near zero. All phase singularities in the VRC either originate or terminate at a voxel with genuinely low signal (such as at the brain boundary) or correspond to the singularity introduced by voxel-wise SVD (green arrow). The latter is successfully removed from the final phase image, resulting in a singularity-free phase image following VRC correction. Empirical evaluations at 3T with a 64-channel receive head coil and at 7T with a 32-channel receive head coil showed that raising the Cholesky operator, $L$, to the power of $n = 1$, always yielded sufficient correlation to generate a VRC with robust coverage in the brain and cervical spinal cord across a range of datasets.

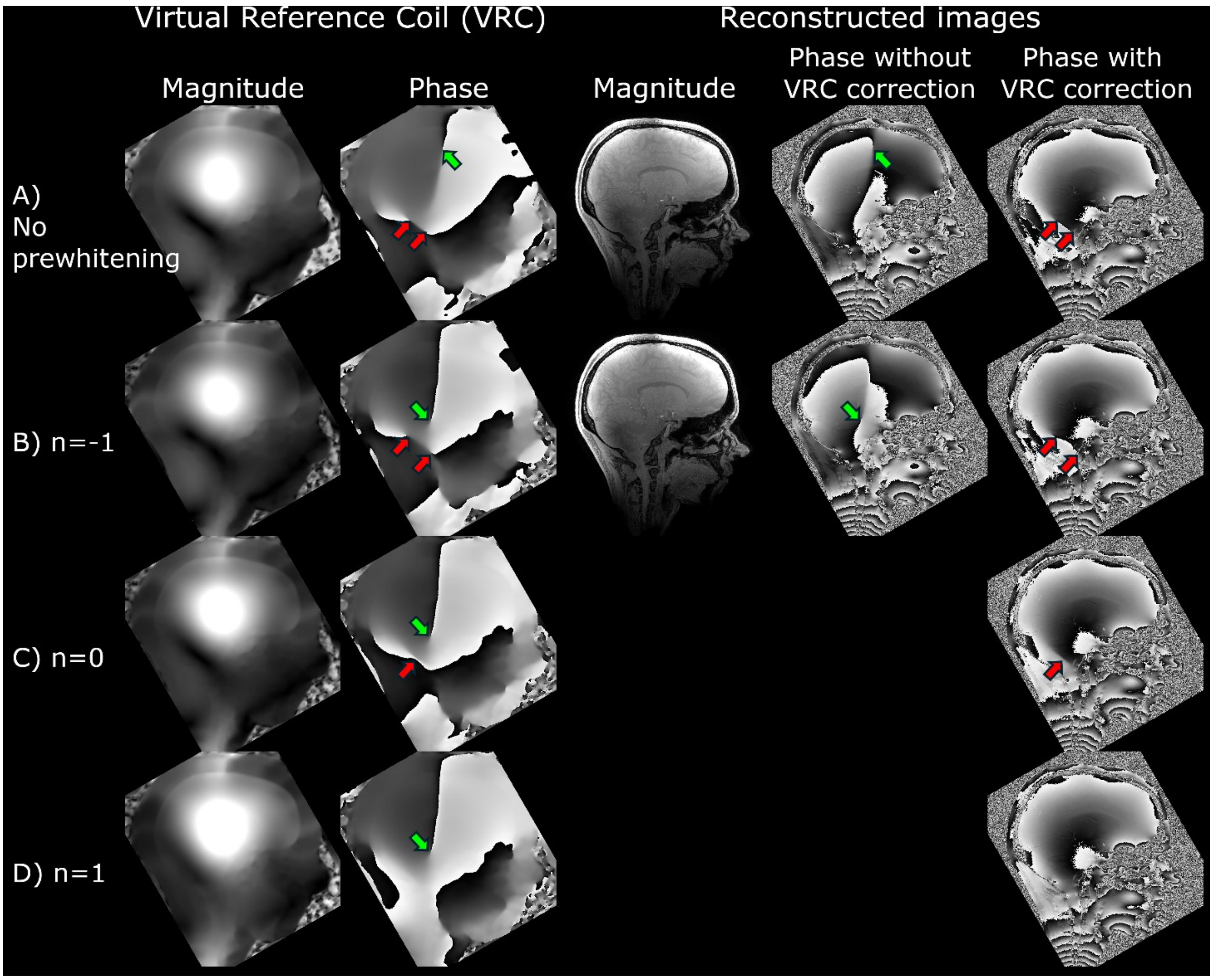

**Figure 2** The effect of pre-whitening and coil correlation on VRC magnitude and phase, and on the final phase image reconstruction illustrated using 7T structural dataset example. MORSE coil sensitivity estimation was performed either without pre-whitening (A) or with (B-D). VRCs were then calculated either using pre-whitened coils sensitivities ($n = -1$), coil sensitivities with inherent correlation ($n = 0$) or with increased correlation power ($n = 1$). Green arrows indicate phase singularities introduced by the voxel-wise SVD used in the coil sensitivity estimation process. Red arrows highlight singularities and phase artefacts arising from VRCs with insufficient coil sensitivity support within the regions of interest (brain and cervical spine). D) represents the final VRC correction solution with $n = 1$, resulting in a final phase image free of phase singularities. The magnitude and phase without VRC correction is identical for B), C) and D). $n > 1$ increased correlation across coils (see Figure 3) but was not required for singularity free phase outputs (data not shown).

We next assessed how increasing the correlation power, $n$, affects the cross-coil correlation and enhances phase coherence between neighbouring coil elements (Figure 3 for a 7T 32-channel head coil). As the correlation power increases from $n = -1$ to 3 the noise covariance matrix shows greater off-diagonal contributions in its magnitude component, alongside increased phase coherence between neighbouring coil elements, reflected in smoother phase patches (see red square in Figure 3 Phase). This stronger phase coherence and magnitude overlap improves VRC support across the entire brain and cervical spinal cord. Analogous results for a 3T 64-channel coil are presented in Supplementary Figure S1.

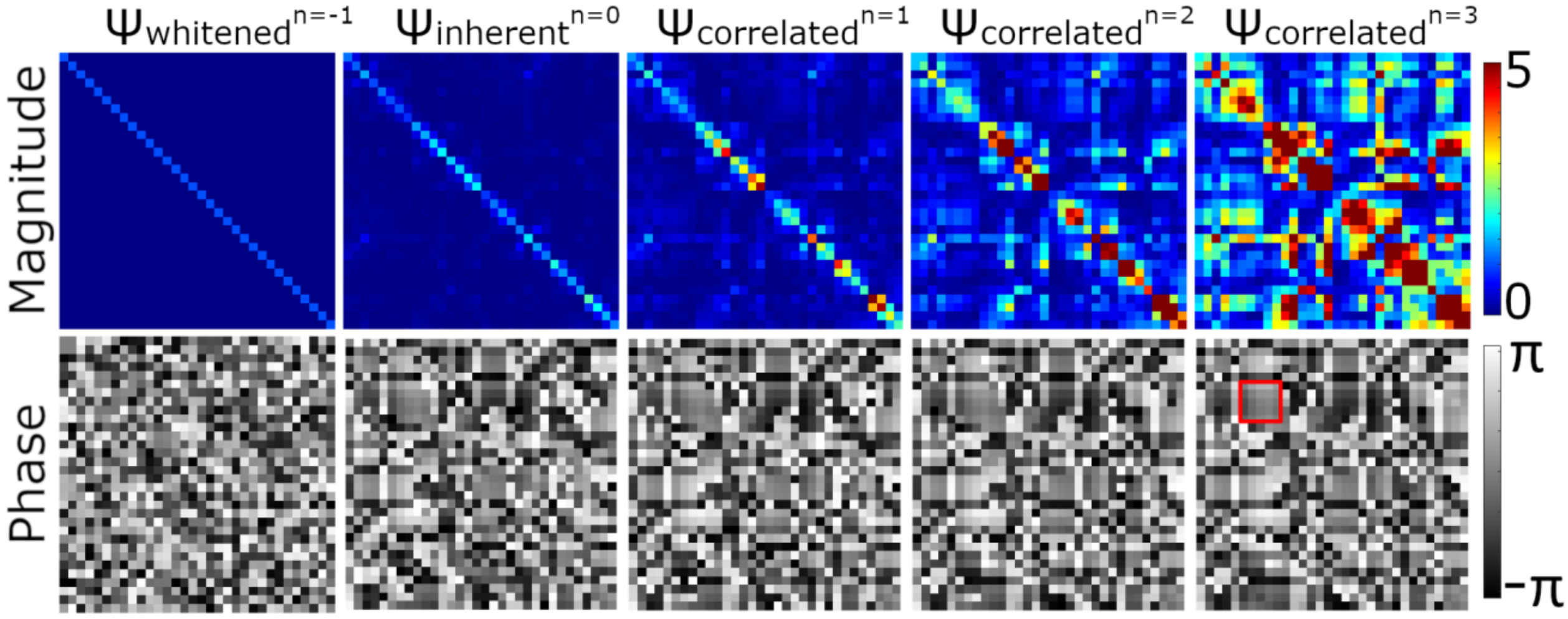

**Figure 3** Noise covariance matrices illustrating the effects of whitening and increasing cross-coil correlation in a 32-channel coil at 7T. $\Psi_{whitened}^{n=-1}$: Noise is independent across coils, resulting in a noise covariance matrix with zero off-diagonal magnitude elements. The phase appears incoherent, exhibiting a characteristic "salt-and-pepper" pattern. $\Psi_{inherent}^{n=0}$: Represents the inherent coil correlations as acquired (i.e., without whitening or artificially increasing correlation). The matrix shows some non-zero off-diagonal magnitude elements and localized patches of phase coherence between neighbouring coil elements. $\Psi_{correlated}^{n=1}$: Represents a level of cross-coil correlation empirically found to provide adequate VRC support. $\Psi_{correlated}^{n=2}$: Off-diagonal magnitude elements are more pronounced, and phase coherence between neighbouring coils increases, forming larger coherent patches. An example of such a patch is highlighted with a red square in the phase component of $\Psi_{correlated}^{n=3}$.

## 4.2 Structural imaging at 7T and 3T - comparison with alternative image reconstruction schemes

### 4.2.1 Image quality assessment

We begin our comparison of MORSE-PI against its predecessor, MORSE. Figure 4 presents exemplary results from a 7T structural MT-weighted scan with four echoes, demonstrating that a clear phase singularity was present in the MORSE reconstruction, which was successfully eliminated in MORSE-PI. Notably, the magnitude images for both reconstructions remained identical.

The phase singularity in the MORSE reconstruction propagated into the CLEAR-SWI results (see Figure 4, red arrows in the 3rd and 4th rows), as Laplacian unwrapping - used in this approach - introduces high frequency phase variations around the singularity. These variations persisted in the phase mask after the SWI high-pass filtering step and could be mistaken for pathology (e.g., potential microbleed or vascular anomaly).

As described in the Methods section, phase singularities can potentially be eliminated using a non-linear complex fit; however, this comes at the cost of global noise amplification, which is readily apparent in Figure 4 in the MORSE-based QSM result relative to that of MORSE-PI. Application of a non-linear complex fit to the MORSE-PI reconstruction provided similarly noisy results (not shown). However, the singularity-free MORSE-PI approach allowed the nonlinear fit to be replaced with the SNR-optimal ROMEO weighted echo averaging, resulting in QSM maps with visibly higher SNR. Both CLEAR-SWI and QSM results based on the MORSE-PI reconstruction showed no residual artifacts and exhibited consistently high image quality.

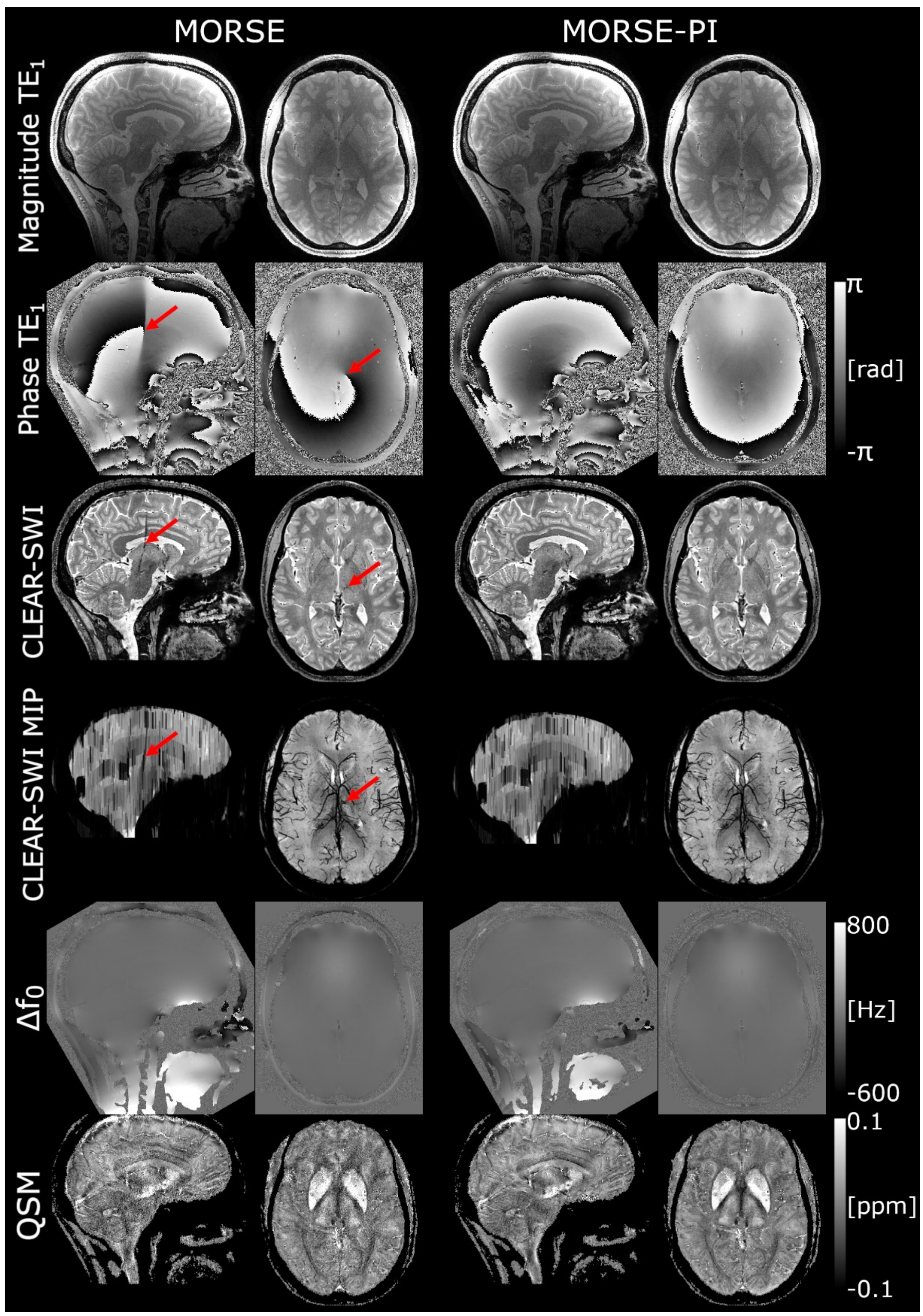

**Figure 4** Comparison of phase reconstruction using the MORSE and MORSE-PI algorithms, along with the corresponding CLEAR-SWI and QSM results. These data are from a 7T MT-weighted scan with four echoes and 0.6mm isotropic resolution. A phase singularity is visible in the MORSE reconstruction (second row, red arrow), which propagates into the CLEAR-SWI result, mimicking pathology in the axial view (e.g., a potential microbleed or venous anomaly). This singularity also necessitates the use of a suboptimal field map calculation method (i.e., a non-linear complex fit), leading to noisier QSM results. In contrast, MORSE-PI phase correction eliminates singularities, enabling $\Delta f_0$ field map calculation using SNR-optimal weighted echo averaging across all echoes. The CLEAR-SWI and QSM from the MORSE-PI reconstruction are artefact-free and exhibit high SNR.

To demonstrate the generalizability of the proposed method and facilitate comparison with the vendor-provided GRAPPA + Adaptive Combine reconstruction, as well as state-of-the-art methods such as ESPIRiT and GRAPPA + ASPIRE, a multi-echo proton density (PD)-weighted scan was performed at 7T using a vendor-supplied sequence.

As shown in Figure 5, the combination of GRAPPA with Adaptive Combine introduced phase singularities in regions where magnitude signal dropouts occurred, forming shapes that could be misinterpreted as pathological features. The resulting phase inconsistencies propagated into the subsequent CLEAR-SWI and QSM reconstructions, as indicated by the red arrows.

While ESPIRiT did not produce magnitude signal voids, it was still affected by phase singularities that propagated into the CLEAR-SWI and QSM reconstructions—particularly when the field map was derived using weighted echo averaging. This issue can be mitigated through non-linear complex fitting during echo combination; however, this approach again comes at the expense of reduced SNR. In line with previous reports[13,37], complex fitting resulted in increased noise in QSM reconstructions with respect to weighted echo averaging across all methods evaluated (compare 4$^{th}$ and 5$^{th}$ rows in Figure 5). GRAPPA + ASPIRE and MORSE-PI yielded qualitatively similar results, both providing singularity-free phase reconstructions and producing high-quality CLEAR-SWI and QSM images.

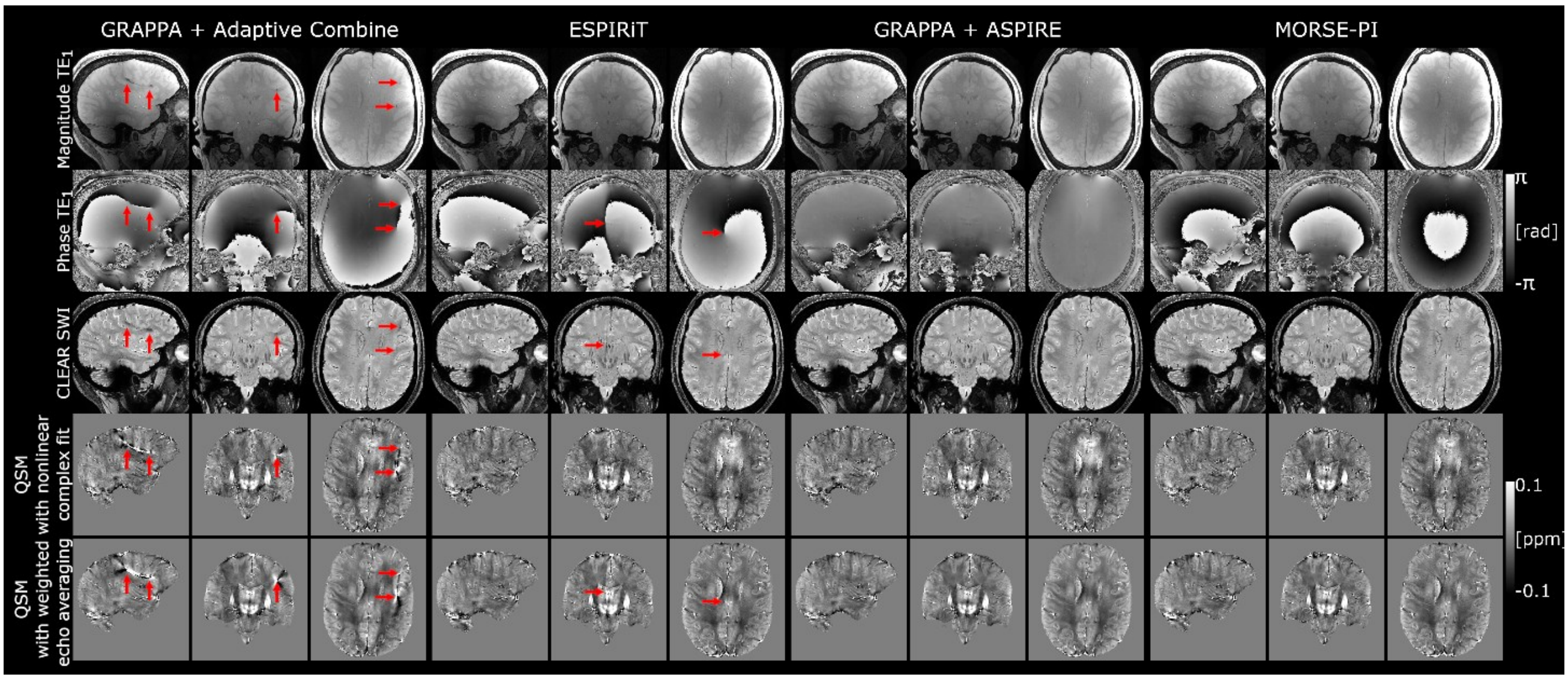

**Figure 5** Comparison of MORSE-PI with GRAPPA + Adaptive Combine, GRAPPA + ASPIRE, and ESPIRiT—methods with potential applications in phase imaging. Red arrows highlight phase singularities in GRAPPA + Adaptive Combine and ESPIRiT, which propagated into the corresponding CLEAR-SWI and QSM images. GRAPPA + Adaptive Combine also showed characteristic magnitude signal voids that mimic pathology. For ESPIRiT, artefacts arising from weighted echo averaging can resemble microbleeds in QSM. Non-linear complex fitting removed some singularities (see ESPIRiT case) but introduced higher noise. GRAPPA + ASPIRE and MORSE-PI yield singularity-free phase images and high-quality CLEAR-SWI and QSM reconstructions.

ASPIRE is a recommended coil combination method for robust phase imaging at ultra-high fields[13], with singularity-free phase maps. However, the overall image quality — particularly regarding noise levels and the presence of fold-over artifacts — highly depends on the quality of the underlying GRAPPA reconstruction of the individual coils. GRAPPA has amplified noise and suffer from residual fold-over artefacts, in both magnitude and phase images at 3T and 7T. These effects are illustrated in two representative 3T cases where the GRAPPA reconstruction resulted in suboptimal image quality: Figure 6 shows fold-over artefacts, while Figure 7 highlights noise amplification associated with a limited field of view and subject-specific head geometry. These artefacts propagate into both the root sum-of-squares (RSS) magnitude images and the ASPIRE-combined phase images.

Figure 6 highlights GRAPPA reconstruction failures in a PD-weighted acquisition with 2×2 acceleration, where aliasing of the ear appeared within the brain (indicated by red, orange, and yellow arrows). These artefacts propagated into the ASPIRE-combined phase images. The green arrow marks an artefact that is effectively masked during phase processing and therefore does not affect the resulting CLEAR-SWI MIP or QSM images. The orange arrow indicates an artefact present only in the first echo, and thus largely suppressed through multi-echo combination. In contrast, the red arrow identifies an artefact present across multiple echoes, which propagates into both CLEAR-SWI and QSM reconstructions, and risks mimicking pathology. In comparison, MORSE-PI reconstruction of the same dataset yielded artefact-free magnitude, phase, CLEAR-SWI, and QSM outputs.

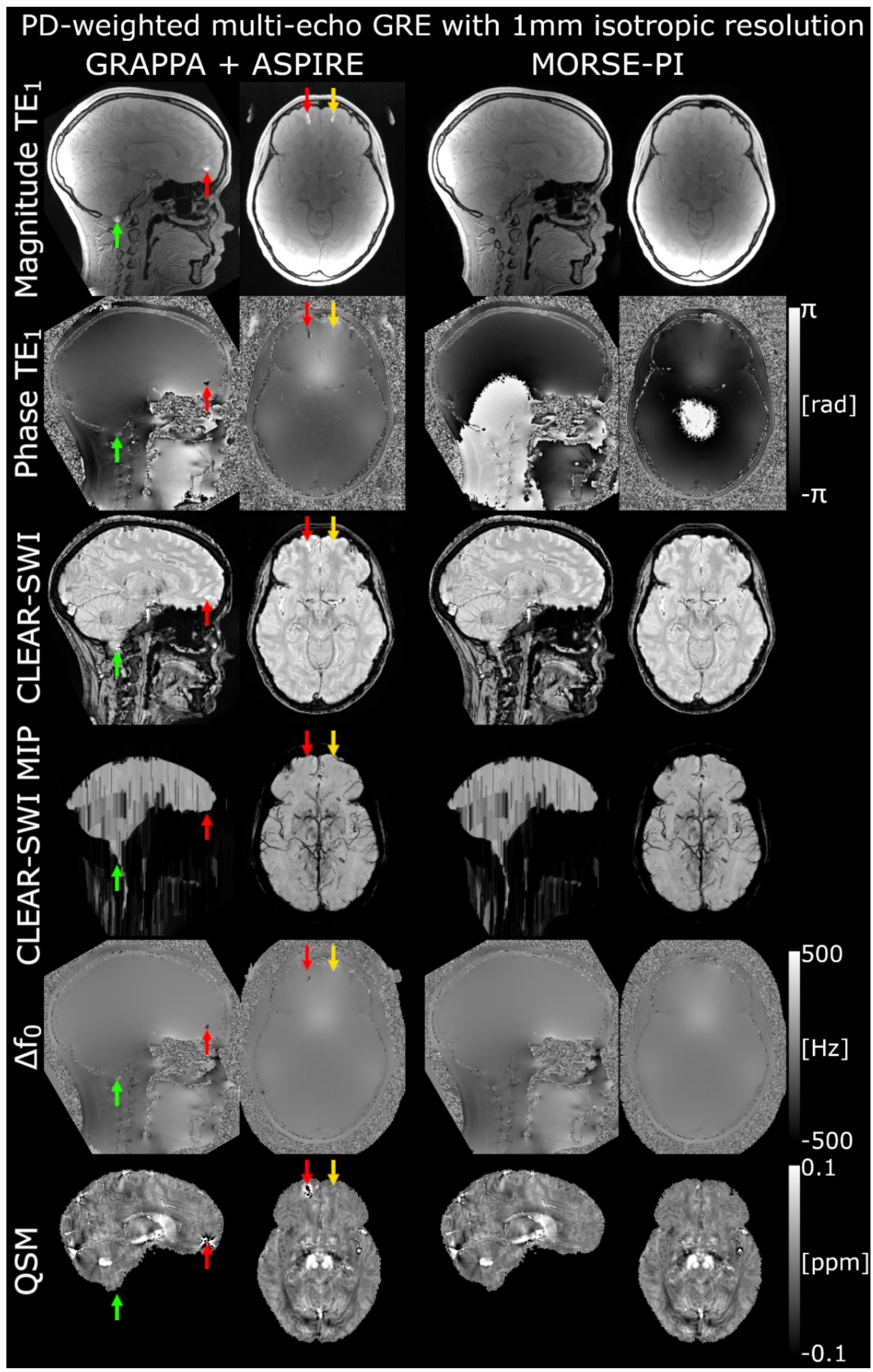

**Figure 6** Comparison of GRAPPA + ASPIRE and MORSE-PI using 3T PD-weighted data (6 echoes, 1 mm isotropic resolution). The GRAPPA reconstruction failed to unfold aliased ear signals to their true location, resulting in residual artefacts (arrows) that propagated into the ASPIRE phase images. These artefacts were only partially removed by echo-weighted $\Delta f_0$ estimation (yellow arrow) or QSM masking (green arrow). A mild hyperintensity appeared in the CLEAR-SWI image, while a prominent streaking artefact was visible in the QSM map (red arrow). In contrast, the MORSE-PI reconstruction was free from these artefacts across all outputs.

Figure 7 presents a 3T example of noise amplification in a GRAPPA reconstruction, a phenomenon frequently observed in participants with larger heads and necks filling or extending beyond the field of view. The GRAPPA + ASPIRE magnitude and phase images exhibited an elevated noise floor, with pronounced noise amplification in posterior regions, including the cerebellum, occipital, and parietal lobes. This noise propagated into the CLEAR-SWI reconstruction, producing artefacts that resembled clusters of microbleeds. As a result, both the CLEAR-SWI and QSM outputs were considered unusable. In contrast, the MORSE-PI reconstruction yielded a substantially lower noise floor and consequently produced high-quality CLEAR-SWI and QSM images.

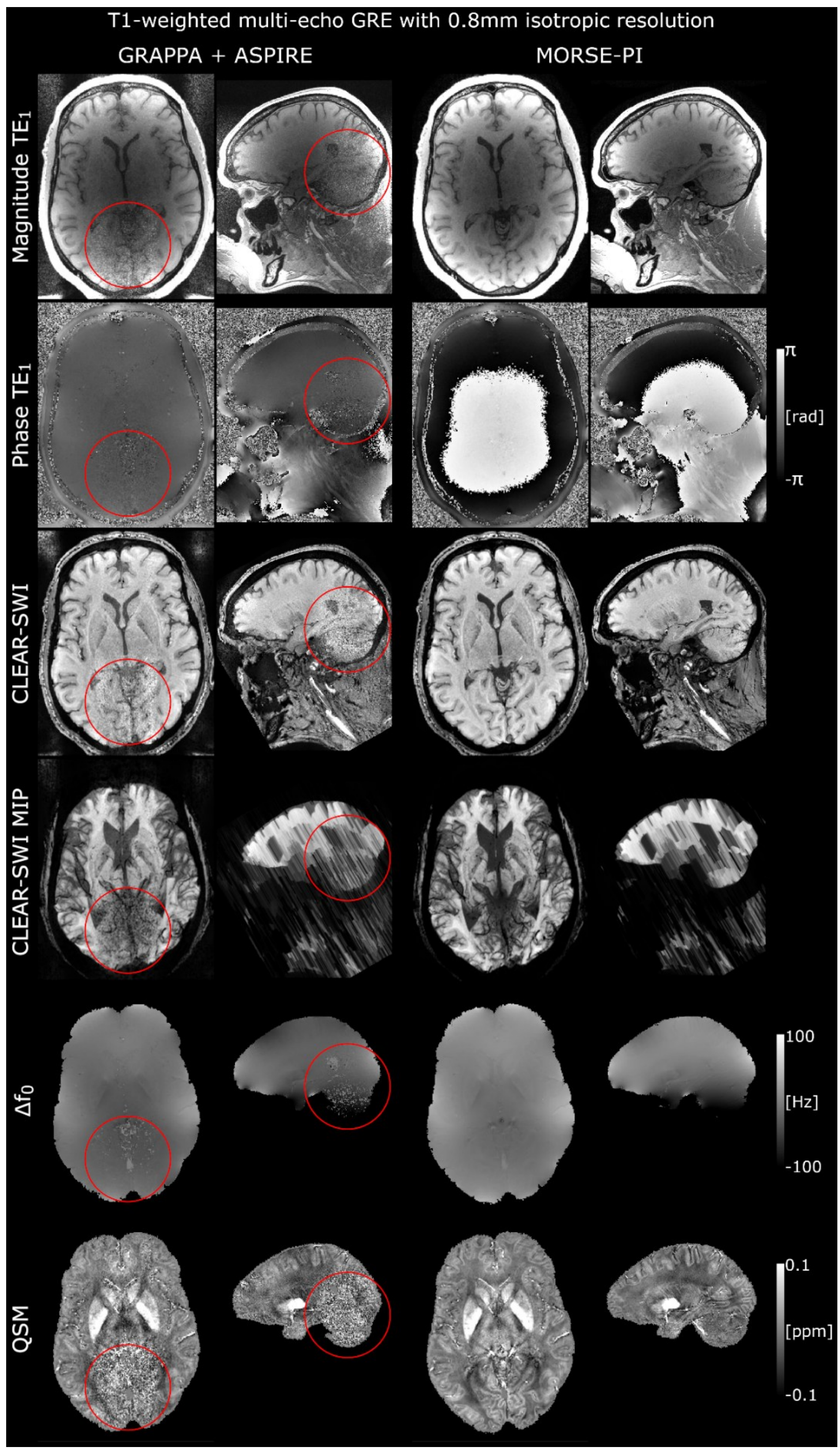

**Figure 7** Example of noise amplification in a GRAPPA reconstruction at 3T, commonly observed in volunteers with large heads and necks filling or extending beyond the field of view. The elevated noise propagated from the GRAPPA magnitude and ASPIRE phase images into the CLEAR-SWI and QSM reconstructions, particularly affecting posterior brain regions and rendering the results unusable. In contrast, the MORSE-PI reconstruction was unaffected by this local noise amplification and had a substantially lower noise floor across the brain, producing high-quality CLEAR-SWI and QSM outputs.

### 4.2.2 Reproducibility of QSM reconstructions

QSM results reconstructed using MORSE-PI exhibited reduced variability compared to those obtained with GRAPPA+ASPIRE, indicative of improved reproducibility. . This was observed consistently in both 3T and 7T datasets, which were free of the substantial noise amplification sometimes present in the GRAPPA reconstructions (c.f. Figure 7). QSM maps were generated from three repeated PD-weighted and T1-weighted acquisitions, all obtained from the same volunteer within a single scan session at each field strength. Notably, substantial inter-scan head motion was observed in both the 3T and 7T sessions, with a maximum rotation of up to 20 degrees about the z-axis. However, these movements were corrected through co-registration of the QSM results, and no phase singularities were present in the data regardless of head position.

Figure 8 shows histograms of QSM SD values within a common brain mask for both reconstruction methods. MORSE-PI exhibited a lower noise floor, as indicated by the blue histogram peaks being closer to zero and showing a smaller right-sided tail compared to the red GRAPPA+ASPIRE distributions. Excluding the scan with the greatest head position deviation from the other two reduced both the QSM SD values and the difference in QSM SD between MORSE-PI and GRAPPA+ASPIRE, although these differences were not fully eliminated (see Supplementary Figure S2). Corresponding QSM mean and SD maps from the three repeats are shown in Supplementary Figure S3. While the mean QSM maps were visually similar between the two reconstruction methods, the globally elevated noise in GRAPPA+ASPIRE was clearly apparent in the SD maps, consistent with the histogram results shown in Figure 8. These findings were consistent across both PD-weighted and T1-weighted datasets at both field strengths.

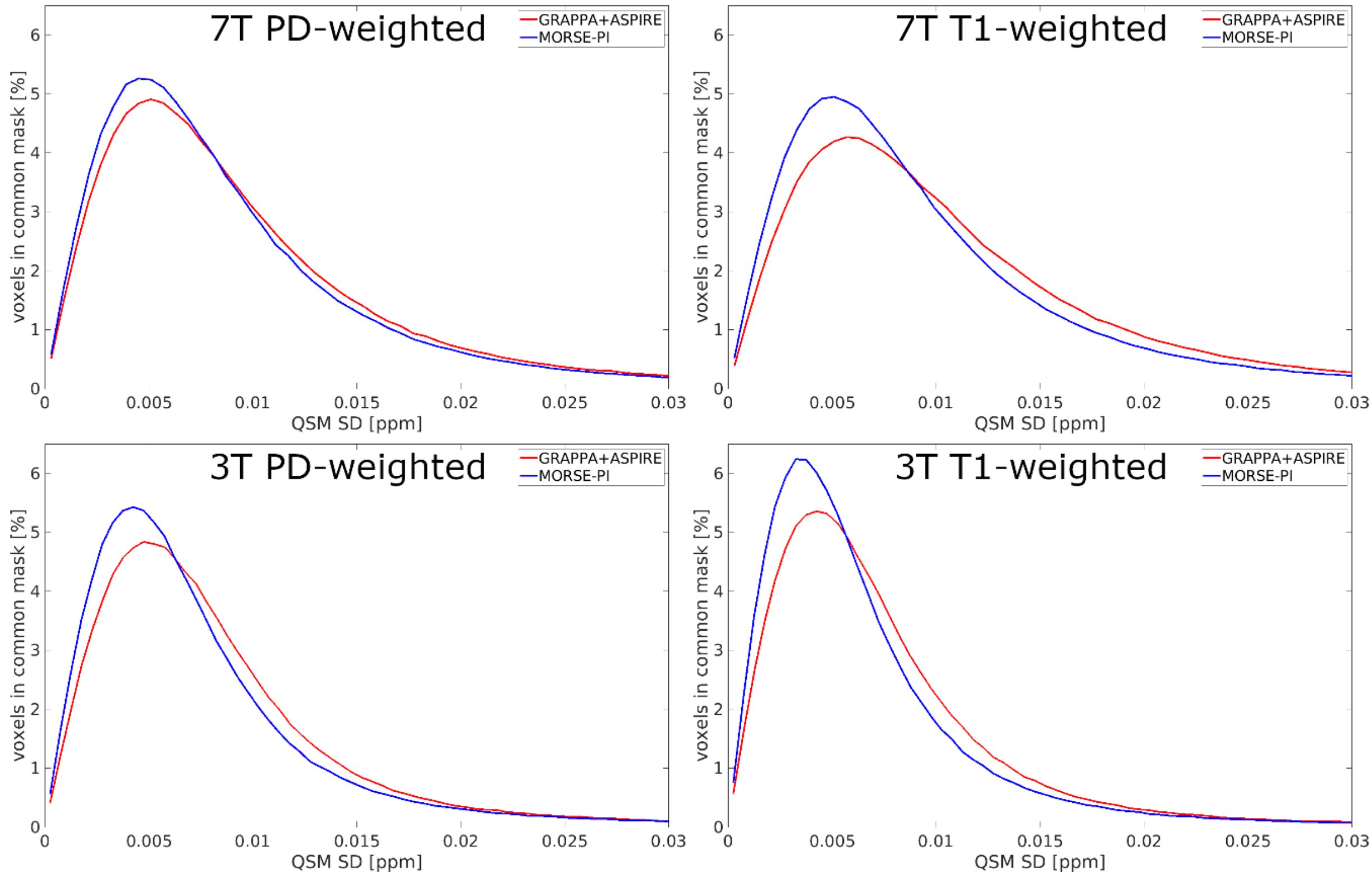

**Figure 8** Reproducibility assessment of GRAPPA+ASPIRE and MORSE-PI image reconstructions from 1mm isotropic GRE scans at 7T and 3T. Reproducibility was quantified by calculating the standard deviation (SD) across three QSMs dervied from three PD-weighted or T1-weighted acquisitions from the same volunteer, performed within a single scan session at each field strength. While the mean QSM maps were visually similar (see Supplementary Figure S2), the GRAPPA+ASPIRE reconstruction exhibited a higher noise floor compared to MORSE-PI, as evidenced by a shift in its histogram peaks toward higher SD values and a higher right tail.

### 4.3 Functional imaging at 7T – evaluation of MORSE-PI phase quality and temporal stability

To evaluate the performance of MORSE-PI on a challenging dataset, we reconstructed images from single-echo, high-resolution (0.8 mm isotropic) 3D EPI acquisitions. These acquisitions present a significant challenge: the ASPIRE and non-linear complex fitting methods cannot be applied as they both require multiple echoes. Figure 9 shows results from both whole-brain and slab-selective single-echo 3D EPI acquisitions[25]. The phase images reconstructed with MORSE-PI were free from singularities, enabling the computation of artefact-free high-SNR QSM maps. This is achieved using rapid 3D EPI scans (e.g., whole-brain EPI: 2 minutes 25 seconds; 4x2 accelerated slab-selective EPI: 4 seconds, acquisition time per volume).

Figure 10 evaluates the temporal stability of both magnitude images and QSMs derived from the slab-selective, high-resolution 3D EPI scan with 50 volumes reconstructed using MORSE-PI. QSM processing for this figure involved PDF background field removal followed by dipole inversion using QSMnet+. The resulting mean magnitude and QSM images exhibited good image quality with no visible artefacts and generally low temporal standard deviation (tSD) values. In the QSM maps, tSD values were noticeably higher in deep grey matter structures and large vessels—regions characterized by high susceptibility and lower signal intensity due to elevated iron content.

To assess the compatibility of MORSE-PI with various background field removal and dipole inversion techniques—as well as with NORDIC denoising[43], which operates on complex-valued data— we computed the temporal mean and SD for selected combinations of post-processing methods (Supplementary Figure S4). While not exhaustive, this analysis demonstrates the flexibility of MORSE-PI in supporting diverse QSM workflows.

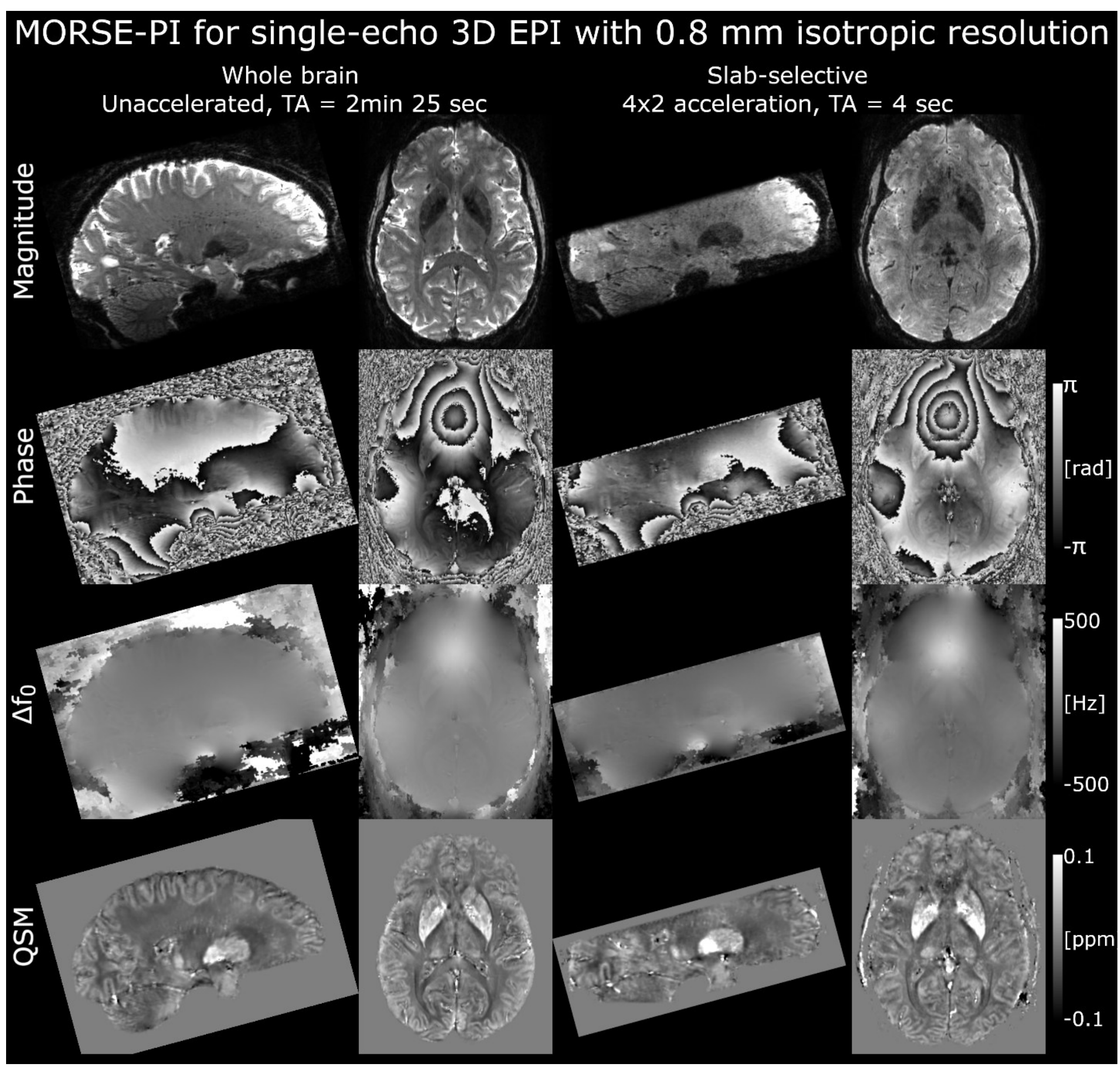

**Figure 9** Exemplar MORSE-PI image reconstruction for high-resolution, single-echo 7T EPI with whole brain and narrow slab coverage[25]. No phase singularities were present in the phase images and QSM maps were of high quality with minimal artefact levels. Good QSM results were obtained from the narrow slab accelerated EPI acquisition with a volume TR of 4 seconds demonstrating the suitability of MORSE-PI for functional QSM applications.

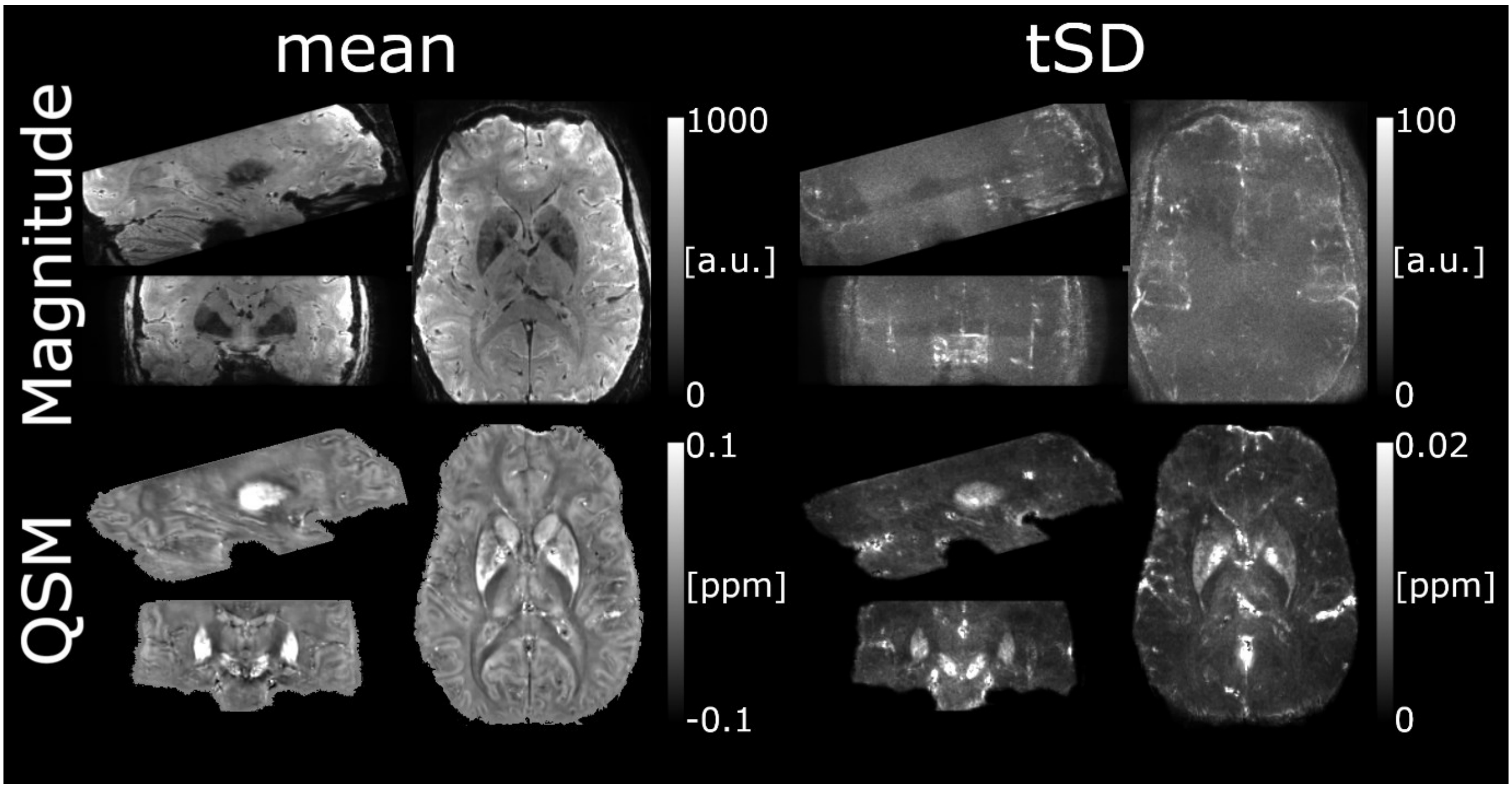

**Figure 10** Demonstration of the temporal stability of MORSE-PI magnitude and phase images in a challenging example of single-echo high-resolution 3D EPI time series (50 volumes). The mean magnitude and QSM images showed no obvious artifacts, and the temporal standard deviation (tSD) was low, except for the QSM tSD in regions with very high magnetic susceptibility, such as deep gray matter regions or large vessels. This QSM result was obtained using PDF and QSMnet+. For other exemplary QSM processing pipelines, as well as the use of NORDIC denoising on MORSE-PI reconstructed images, see Supplementary Figure S4.

# 5. Discussion

The MORSE framework is a regularised SENSE-based image reconstruction method with computational requirements that are commensurate with online deployment that produces magnitude images with low artefact levels even when under-sampled. The extension presented here, called MORSE-PI, generates high-SNR phase images that are free from singularities in the evaluated high- and ultra-high-field scenarios, making it well-suited for phase-based imaging techniques such as QSM, SWI, MR thermometry, distortion correction, or transmit field mapping.

MORSE-PI builds on the previously described concept of a Virtual Reference Coil (VRC) with scalar phase matching at a seed voxel[20]. However, it introduces key improvements. In the original method, the VRC is derived from coil-wise complex images. This can lead to regions without VRC support—such as near ear canals, sinuses, or large vessels—resulting in phase singularities[11]. In contrast, MORSE-PI constructs the VRC using coil-wise sensitivity maps estimated with the robust MORSE method. This provides better spatial support in these problematic regions.

To further strengthen the VRC robustness at ultra-high field strengths and for coil arrays with many channels (≥32), we propose leveraging the coil covariance matrix to enhance the spatial correlation between neighbouring coil sensitivities prior to VRC calculation. The coil covariance matrix is obtained from rapid noise-only pre-scans and therefore does not increase scan time. This matrix is typically used for noise whitening before image reconstruction— boosting SNR and producing steeper coil sensitivity profiles that improve data unfolding. However, these steeper profiles are suboptimal for generating a VRC with reliable spatial support across the full field of view. Therefore, MORSE-PI applies the inverse of the whitening process to instead increase cross-coil correlation and spatial overlap. This ensures that the VRC, calculated as the complex sum of coil sensitivities, has consistent support across the entire field of view.

The level of spatial correlation is controlled by repeatedly applying the inverse of the pre-whitening matrix (i.e. $n = 1,2,3, \dots$ in Equation (5)). In practice, applying the inverse once to unwhitened data or twice to whitened data (i.e., $n = 1$) has proved sufficient. This approach performs well across different acquisition types, including EPI and GRE, at both 3T and 7T, using 64- and 32-channel head coils. Higher values of $n$ may be required at higher field strengths or with coil arrays composed of smaller elements. However, in regions where there is inherently no coil sensitivity across any coil element, the absence of magnitude signal, and the resulting phase singularities, will persist. Importantly, this correlation enhancement step is only applied during VRC construction—it does not affect the image unfolding process and is compatible with both whitened and unwhitened data. The method requires only a noise calibration scan, which is routinely acquired in many MRI protocols.

To ensure accurate scalar phase matching between coil sensitivities before the complex summation, we select a seed voxel based on the weighted centroid (using regionprops3 MATLAB function) of the magnitude root-sum-of-squares image from the calibration data used to estimate the sensitivities. This reliably identifies a voxel where all coils exhibit non-zero sensitivity, regardless of head position within the coil array (as demonstrated here in 1mm isotropic resolution GRE data that had up to 20 degrees rotation about the z-axis).

We compared MORSE-PI to state-of-the-art phase image reconstruction methods and demonstrated that it produced superior phase image quality while maintaining the excellent magnitude image quality of the original MORSE method. In our prior work, MORSE outperformed GRAPPA, ESPIRiT, ENLIVE[44], and LORAKS[45] when reconstructing high-resolution magnitude data from datasets similar to those used in this manuscript. MORSE-PI extends this advantage to phase images. The ASPIRE method, which operates on individual coil images following GRAPPA reconstruction, propagates any noise amplification or fold-over artefacts present in the separate coil data to the final image. We showed that MORSE-PI outperforms the GRAPPA + ASPIRE combination, particularly in QSM applications, which are highly sensitive to noise due to the ill-posed nature of dipole inversion.

SVD coil compression, when applied before ESPIRiT or GRAPPA + Adaptive Combine reconstruction, can potentially mitigate phase singularities[42]. The dominant singular vector (i.e., the first compressed coil with the largest signal) often has a smoother phase profile and may therefore act as a VRC. However, at higher field strengths and with coil arrays consisting of smaller elements, the spatial profiles of coil sensitivities become steeper[46]. In such cases, the dominant singular vector may exhibit a more heterogeneous phase profile, including singularities, which then propagate to the final phase image. Supplementary Figure S5 shows examples at 3T (with SVD + ESPIRiT) and 7T (with SVD + GRAPPA + Adaptive Combine, a vendor-provided option described in the QSM Consensus Organization Committee publication Supporting Information III[13]), both of which fail to produce singularity-free phase images.

MORSE-PI is a highly flexible reconstruction method that delivers high-quality magnitude and phase images at both 3T and 7T for various acquisitions. These include multi- and single-echo 3D GRE and EPI, as well as phase-based $B_1$ mapping (see Supplementary Figure S6)[47]. MORSE-PI does not require additional reference body coil[39] or multi-echo[48,49] reference scans. Single-echo 3D EPI at 7T is a particularly challenging scenario, where existing approaches typically require additional reference scans to reliably eliminate phase singularities[49]. To further assess the robustness of MORSE-PI under these conditions, we evaluated its performance across seven additional variations of the 3D EPI protocol described in Table 1[25,50], each acquired in a different volunteer. These variations included both segmented and non-segmented readouts, Partial Fourier applied at either the beginning or end of k-space plane, 1-3-3-1 or 1-2-1 binomial excitations, a range of field-of-view configurations (from narrow slab to full-brain coverage), and the presence of intentional head motion (rotations from +6° to −4° relative to separately acquired fully sampled calibration data). In all cases, MORSE-PI produced reconstructions free of phase singularities (data not shown). Although developed for neuroimaging, MORSE-PI is readily applicable to other anatomical regions. Supplementary Figure S7 shows exemplar QSM results of the liver at 3T[51] and the knee at 7T[52] derived from MORSE-PI reconstructed images.

MORSE-PI is implemented using the vendor-agnostic Gadgetron framework encapsulated in Docker containers, enabling rapid online reconstruction. For example, a 0.6 mm isotropic 3D GRE acquisition with a 9-minute 10-second scan time is reconstructed approximately 60 seconds after the scan ends on a Dell Precision 7920 workstation with an Intel Xeon Gold 6230R 2.1GHz processor and 768 GB of RAM. For long EPI time series, image reconstruction completes almost instantaneously after scanning completes, as the most computationally expensive step—coil sensitivity estimation—is completed at the outset of the acquisition.

The high-SNR, artefact-free phase images produced by MORSE-PI provide an excellent foundation for downstream processing techniques such as SWI, QSM, fQSM or STI. MORSE-PI is compatible with a wide range of $B_0$ map estimation techniques (see Figure 5), background field removal methods, dipole inversion algorithms, and denoising pipelines (see Supplementary Figure S4). While non-linear complex fitting can be used to derive $B_0$ maps from multi-echo phase images corrupted by singularities (as in the ESPIRiT reconstruction in Figure 5), this yields noisier results than weighted echo averaging in agreement with prior findings in the literature[13,37]. In contrast, the absence of phase singularities in MORSE-PI enables optimal echo combination via weighted averaging, leading to improved QSM quality and robustness. Moreover, MORSE-PI's versatility across single- and multi-echo GRE and EPI acquisitions, at various field strengths and with different multi-channel coils, makes it particularly well-suited for multi-centre and cross-field-strength studies, where consistency in image reconstruction is desirable.

# 6. Conclusions

MORSE-PI flexibly provides high SNR, fold-over-free and singularity-free phase images, as demonstrated for single-echo and multi-echo structural GRE and functional EPI scans. MORSE-PI naturally lends itself to imaging techniques, such as structural and functional QSM, that necessitate high-quality phase images.

## Code and Data Availability Statement

The open-source code, as well as scripts necessary to compile a Gadgetron image and create a Docker container encapsulating the MORSE-PI reconstruction, are available on:

https://github.com/fil-physics/gadgetron-matlab.

Example 3T and 7T multi-echo GRE raw k-space datasets are available on:
https://zenodo.org/records/13737462

In-house developed GRE sequence for Siemens 3T Prisma system:
https://xip.uclb.com/product/mri-pulse-sequence-for-a-multi-echo-spoiled-gradient-echo

In-house developed $B_1$ sequence for Siemens 7T Terra system:
https://xip.uclb.com/product/mri-pulse-sequence-for-b1

Our CLEAR SWI and QSM processing pipelines for MORSE-PI are available on:
https://github.com/fil-physics/MPM_QSM

## Authors Contributions

Dymerska: Study conception and design, Code contributions, Acquisition design, Acquisition of data, Analysis and interpretation, Drafting of manuscript, Critical revision, Josephs: Code contributions, Analysis and interpretation, James: Code contributions, Acquisition of data, Malekian: Acquisition design, Critical revision, Graedel: Acquisition design, Acquisition of data, Critical revision, Callaghan: Study conception and design, Code contributions, Acquisition design, Acquisition of data, Critical revision.

## Compliance with ethical standars

**Funding:** This research was funded in whole, or in part, by the Wellcome Trust [203147/Z/16/Z and 226793/Z/22/Z].

**Conflict of interest:** The authors have no competing interests to declare that are relevant to the content of this article.

**Ethical approval:** All procedures performed in studies involving human participants were in accordance with the ethical standards of the institutional and/or national research committee and with the 1964 Helsinki declaration and its later amendments or comparable ethical standards.

**Informed consent:** Informed consent was obtained from all individual participants included in the study.

## Supplementary material

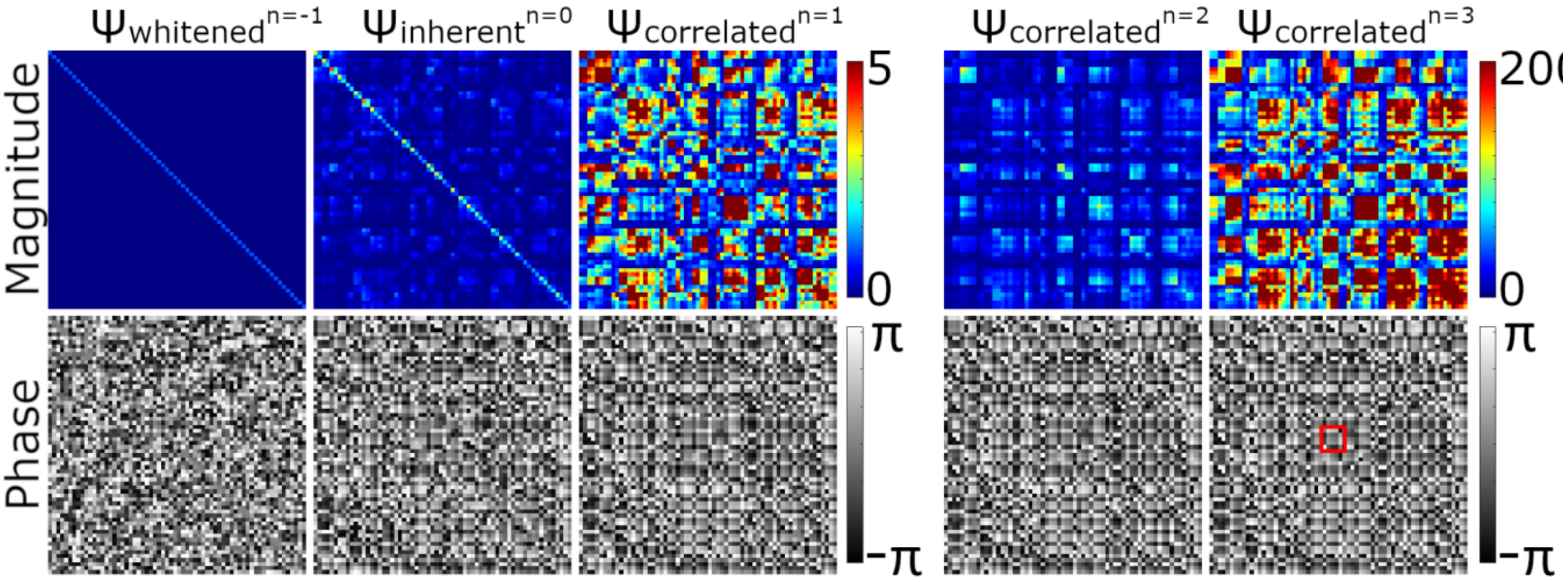

**Figure S1** Noise covariance matrices illustrating the effects of whitening or increasing cross-channel correlation in a 64-channel coil at 3T. $\Psi_{whitened}{}^{n=-1}$: Noise is independent across channels, resulting in a noise covariance matrix with zero off-diagonal magnitude elements. The phase appears incoherent, exhibiting a characteristic "salt-and-pepper" pattern. $\Psi_{inherent}{}^{n=0}$: Represents the inherent channel correlations as acquired (i.e., without whitening or increasing correlation). The matrix shows more non-zero off-diagonal magnitude elements than at 7T for 32-channel coil (compare with Figure 3 in main manuscript). $\Psi_{correlated}{}^{n=1}$: Represents a level of cross-coil correlation empirically found to provide adequate VRC support. $\Psi_{correlated}{}^{n=2}$: Off-diagonal magnitude elements are more pronounced, and phase coherence between neighbouring channels increases, forming more coherent patches. An example of such a patch is highlighted with a red square in the phase component of $\Psi_{correlated}{}^{n=3}$.

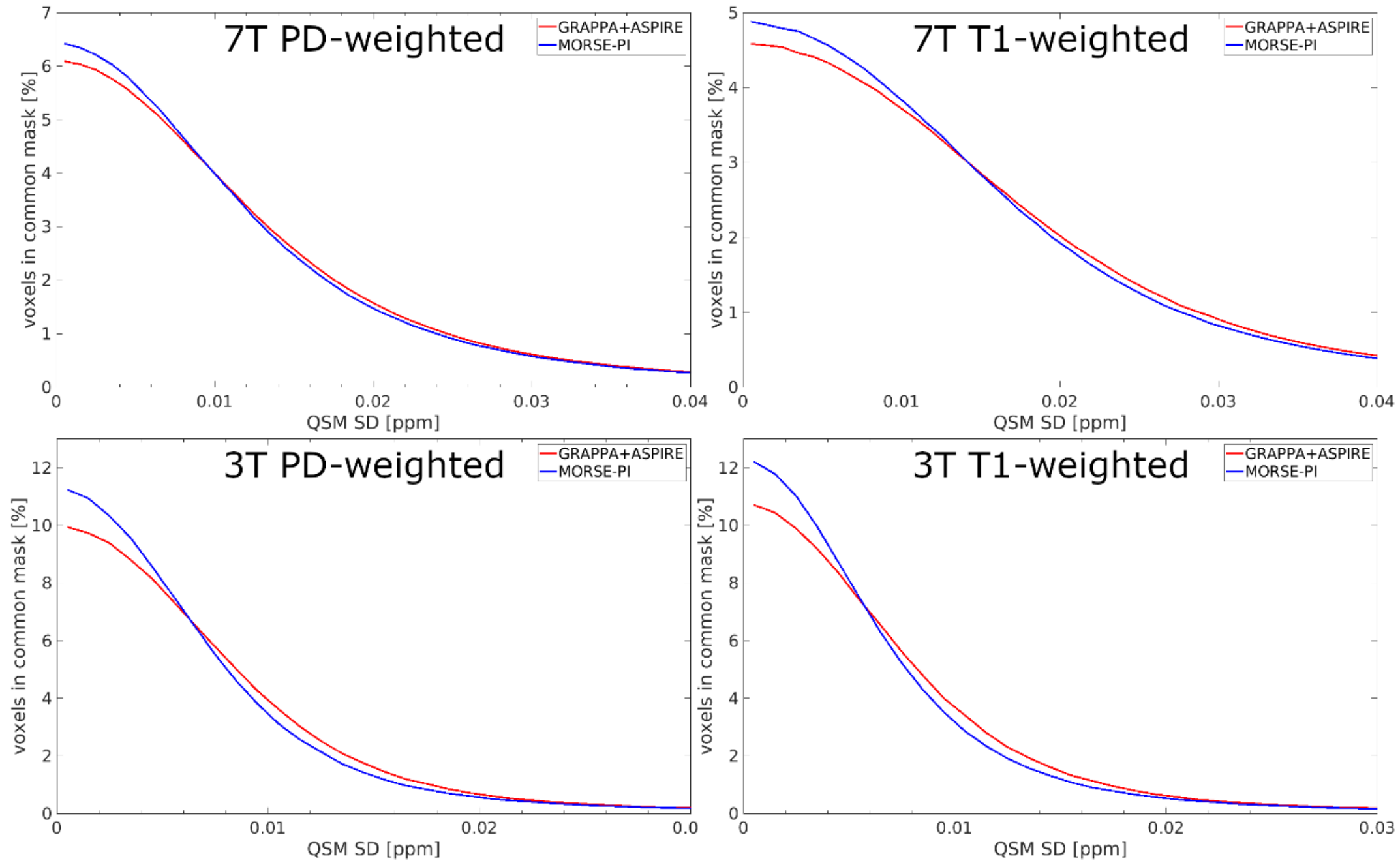

**Figure S2** Reproducibility assessment at 7T and 3T for GRAPPA+ASPIRE and MORSE-PI image reconstructions of PD-weighted and T1-weighted scans. Reproducibility was quantified using the standard deviation (SD) across two QSMs based on two out of three acquisitions from the same volunteer, acquired within the same scan session at each field strength. The two scans with the least inter-scan motion were selected for analysis. GRAPPA+ASPIRE reconstruction exhibited a higher noise floor than MORSE-PI, as evidenced by lower peak at zero and a more extended right tail in the distribution of QSM SD values.

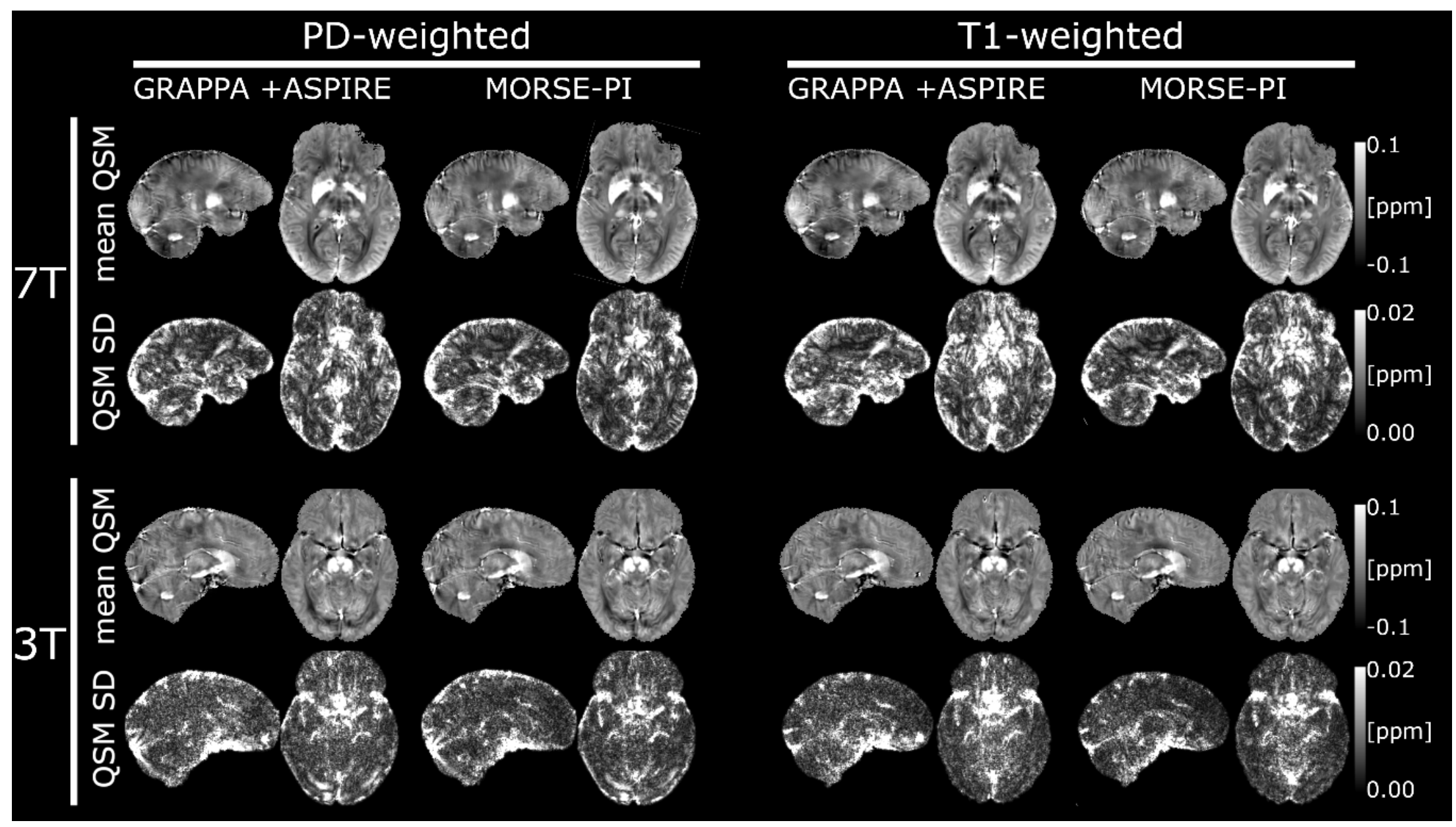

**Figure S3** Mean and standard deviation (SD) maps computed from three QSM reconstructions based on repeated PD-weighted or T1-weighted acquisitions of the same volunteer. All scans were performed within a single scan session at each field strength. While the mean QSM maps appear visually similar between GRAPPA+ASPIRE and MORSE-PI for a given contrast and field strength, the QSM SD maps reveal globally increased variability in the GRAPPA+ASPIRE reconstructions. This is consistent with Figure 8 based on the same data, where GRAPPA+ASPIRE histogram peaks are shifted towards higher QSM SD values and have higher right tail than MORSE-PI.

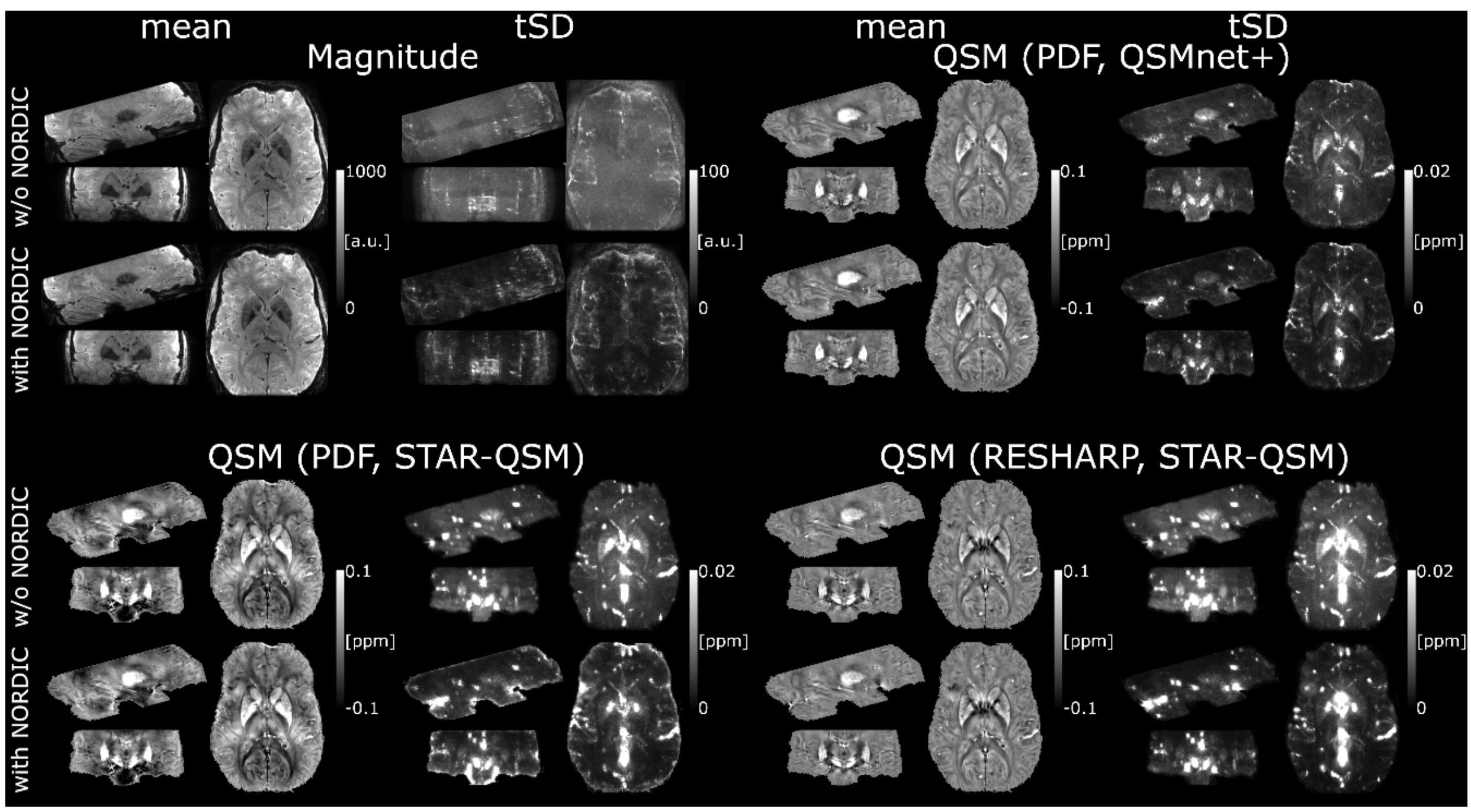

**Figure S4** Mean and temporal standard deviation (tSD) maps from MORSE-PI reconstruction of a 7T accelerated single-echo 3D EPI time series, processed using different QSM pipelines. These include two background field removal methods (PDF and RESHARP), two dipole inversion techniques (QSMnet+ and STAR-QSM), and optional NORDIC denoising. While not exhaustive, this demonstrates the flexibility of MORSE-PI in supporting diverse QSM workflows. The lowest tSD was achieved using the PDF + QSMnet+ pipeline with NORDIC denoising. Corresponding MORSE-PI mean and SD magnitude images are included for reference. This figure expands on the results presented in Figure 10 of the main manuscript.

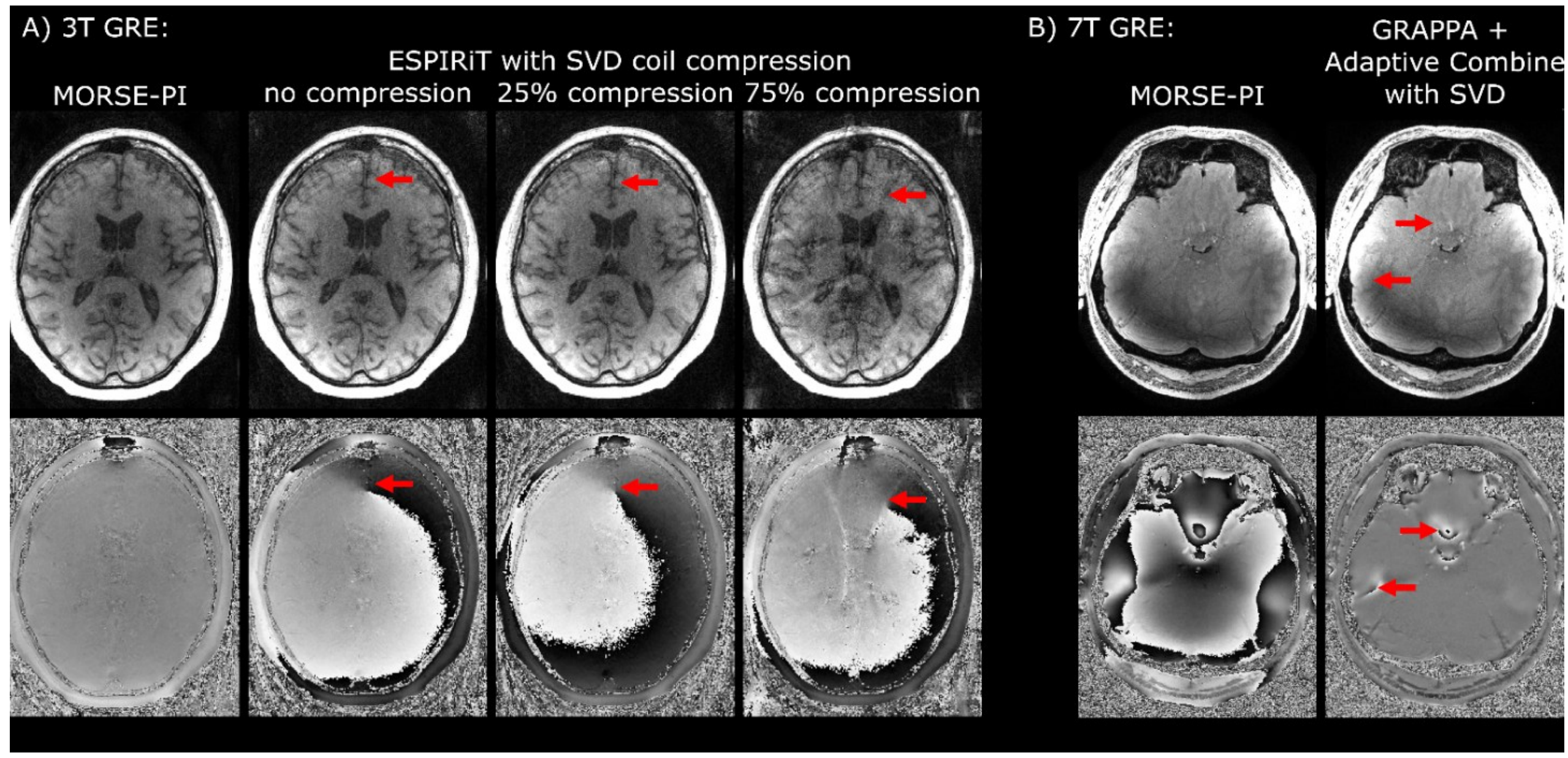

**Figure S5** Evaluation of SVD-based coil compression for phase singularity removal prior to image reconstruction. (A) ESPIRiT reconstruction at 3T with SVD coil compression. (B) GRAPPA + Adaptive Combine reconstruction at 7T, using a vendor-provided retrospective reconstruction that incorporates SVD. In both cases, phase singularities persist in regions of high signal magnitude (red arrows), demonstrating that SVD does not eliminate them. Light (25%) and heavy (75%) coil compression had only minor effects on the location of open-ended fringe lines, although heavy compression led to noticeable image degradation. Although GRAPPA + Adaptive Combine with SVD eliminated signal voids in the magnitude images, phase singularities persisted—typically near the brain boundaries but still within areas of high magnitude signal. MORSE-PI reconstruction is also provided for (A) and (B) as reference.

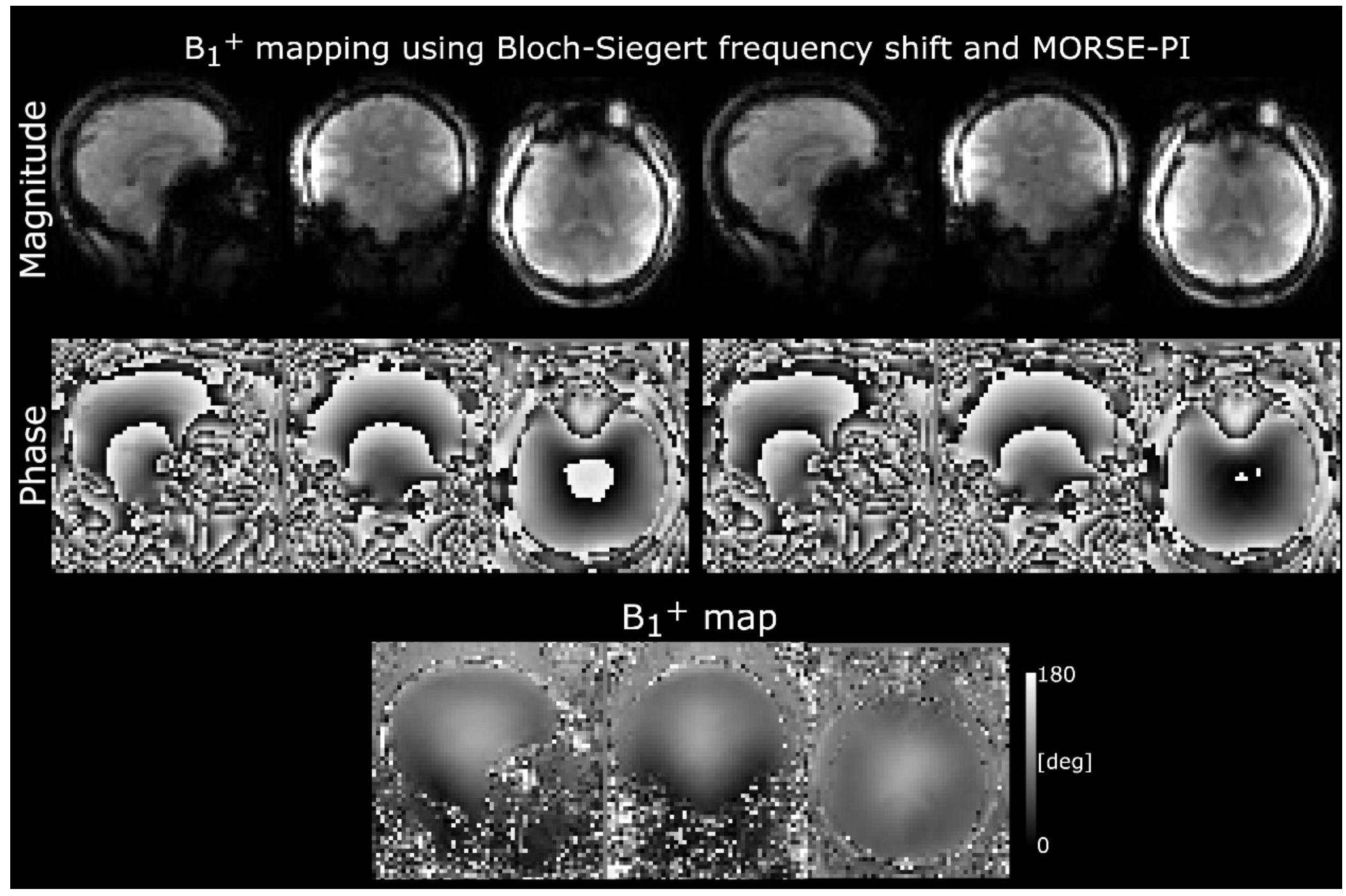

**Figure S6** Example of MORSE-PI reconstruction of a 7T B1⁺ map using the Bloch-Siegert frequency shift method. The two phase images acquired with opposite off-resonance RF pulses are free of phase singularities. The resulting B1⁺ map, computed from the phase difference between these acquisitions, is of high quality.

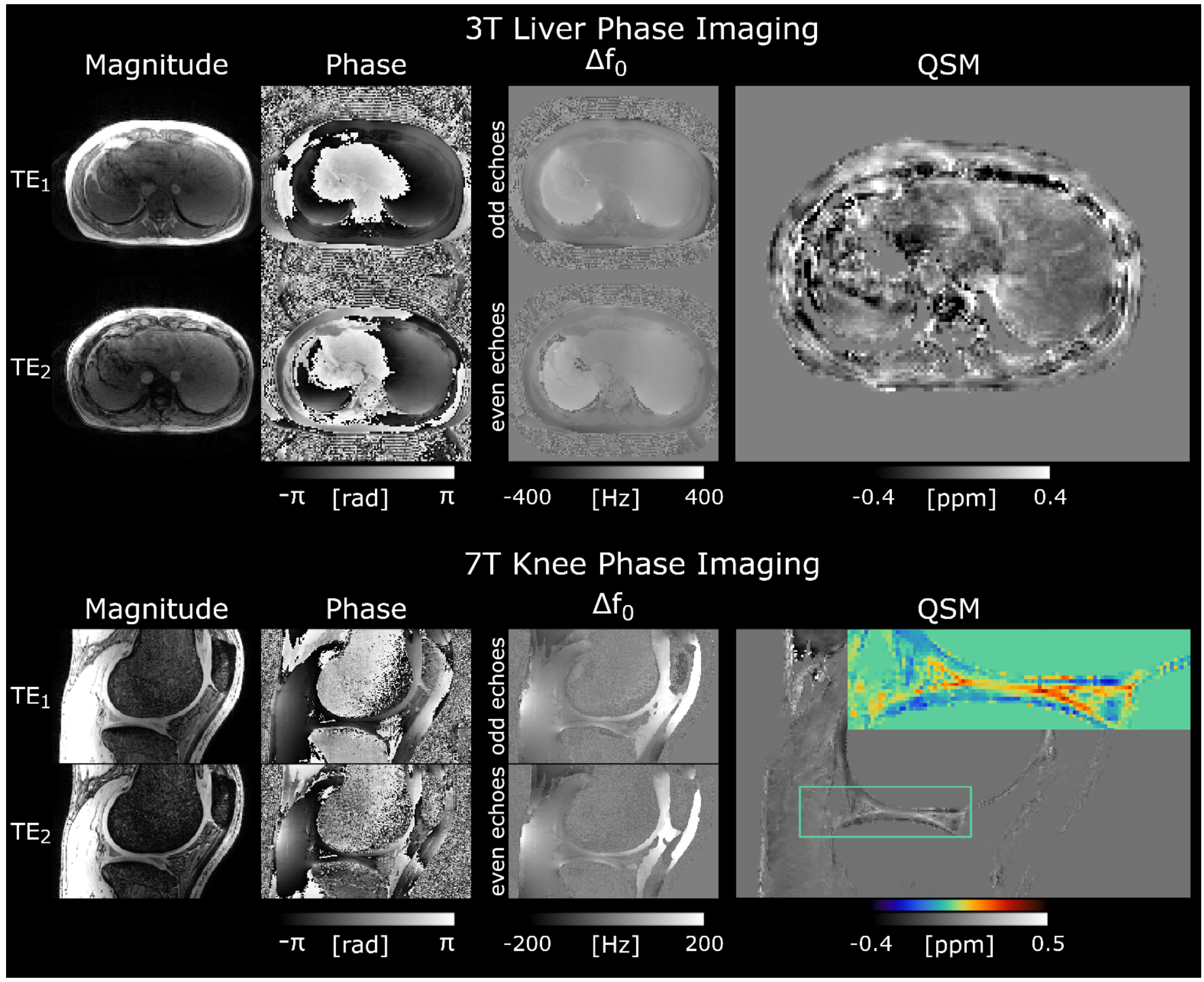

**Figure S7** MORSE-PI reconstructions of multi-echo gradient-echo images acquired in the liver (3 T) and knee (7 T). For liver imaging (2.5 × 2.5 × 3.0 $mm^3$ resolution; acceleration 2 (PE) × 2 (3D)), echo times were selected to sample alternating in-phase and out-of-phase water–fat signals (six echoes; $TE_1$ = 2.32 ms, ΔTE = 1.16 ms). For knee imaging (0.6 mm isotropic resolution; acceleration 2 (PE) × 2 (3D)), echo times were chosen to cover a range suitable for cartilage and meniscus T2* mapping (eight echoes; $TE_1$ = 2.30 ms, ΔTE = 2.38 ms). Frequency shift maps ($\Delta f_0$) were estimated separately from odd and even echoes to account for the substantial water–fat chemical shift, and subsequently averaged prior to background field removal (PDF) and dipole inversion (STAR-QSM). High-quality QSM reconstructions were obtained, particularly in the knee, where subtle cartilage layering is clearly visible (see coloured QSM zoom panel).